\documentclass[twocolumn]{aastex62}

\usepackage{amsmath}
\usepackage{amstext}
\usepackage{apjfonts}
\usepackage{graphicx}

\usepackage{amssymb}
\usepackage{latexsym}
\usepackage{longtable}

\defcitealias{Narayan16}{N16}
\defcitealias{Calamida18}{C18}

\shorttitle{Photometry and spectroscopy of faint candidate standard white dwarfs} 
\shortauthors{Calamida et al.}

\begin{document}

\title{Photometry and spectroscopy of faint candidate spectrophotometric standard DA white dwarfs}

\correspondingauthor{Annalisa Calamida}
\email{calamida@stsci.edu}

\author{Annalisa Calamida}
\affiliation{National Optical Astronomy Observatory - AURA, 950 N Cherry Ave, Tucson, AZ, 85719, USA}
\affiliation{Space Telescope Science Institute - AURA, 3700 San Martin Drive, Baltimore, MD 21218, USA}

\author{Thomas Matheson}
\affiliation{National Optical Astronomy Observatory - AURA, 950 N Cherry Ave, Tucson, AZ, 85719, USA}

\author{Abhijit Saha}
\affiliation{National Optical Astronomy Observatory - AURA, 950 N Cherry Ave, Tucson, AZ, 85719, USA}

\author{Edward Olszewski}
\affiliation{Steward Observatory, University of Arizona, 933 N Cherry Ave, Tucson, AZ, 85719, USA}

\author{Gautham Narayan}
\affiliation{National Optical Astronomy Observatory - AURA, 950 N Cherry Ave, Tucson, AZ, 85719, USA}
\affiliation{Space Telescope Science Institute - AURA, 3700 San Martin Drive, Baltimore, MD 21218, USA}

\author{Jenna Claver}
\affiliation{National Optical Astronomy Observatory - AURA, 950 N Cherry Ave, Tucson, AZ, 85719, USA}

\author{Clare Shanahan}
\affiliation{Space Telescope Science Institute - AURA, 3700 San Martin Drive, Baltimore, MD 21218, USA}

\author{Jay Holberg}
\affiliation{Lunar and Planetary Laboratory,  University of Arizona, 1629 E University Blvd, Tucson, AZ 85721, USA}

\author{Tim Axelrod}
\affiliation{Steward Observatory, University of Arizona, 933 N Cherry Ave, Tucson, AZ, 85719, USA}

\author{Ralph Bohlin}
\affiliation{Space Telescope Science Institute - AURA, 3700 San Martin Drive, Baltimore, MD 21218, USA}

\author{Christopher W. Stubbs}
\affiliation{Harvard University, 17 Oxford Street, Cambridge MA 02138, USA}

\author{Susana Deustua}
\affiliation{Space Telescope Science Institute - AURA, 3700 San Martin Drive, Baltimore, MD 21218, USA}

\author{Ivan Hubeny}
\affiliation{Steward Observatory, University of Arizona, 933 N Cherry Ave, Tucson, AZ, 85719, USA}

\author{John Mackenty}
\affiliation{Space Telescope Science Institute - AURA, 3700 San Martin Drive, Baltimore, MD 21218, USA}

\author{Sean Points}
\affiliation{Cerro Tololo Inter-American Observatory, Casilla 603, La Serena, Chile}

\author{Armin Rest}
\affiliation{Space Telescope Science Institute - AURA, 3700 San Martin Drive, Baltimore, MD 21218, USA}
\affiliation{Department of Physics and Astronomy, Johns Hopkins University, Baltimore, MD 21218, USA}

\author{Elena Sabbi}
\affiliation{Space Telescope Science Institute - AURA, 3700 San Martin Drive, Baltimore, MD 21218, USA}


\begin{abstract}

We present precise photometry and spectroscopy for 23 candidate spectrophotometric standard white dwarfs. 
The selected stars are distributed in the Northern hemisphere and around the celestial equator
and are all fainter than $r \sim$ 16.5 mag. This network of stars, when established as standards, together with the three 
\emph{Hubble Space Telescope} primary CALSPEC white dwarfs, will provide a set of spectrophotometric standards 
to directly calibrate data products to better than 1\%.
These new faint standard white dwarfs will have enough signal-to-noise ratio in future deep photometric 
surveys and facilities to be measured accurately while still avoiding saturation in such surveys. They will also fall within the dynamic range of large telescopes and their instruments for the foreseeable future.  This paper discusses the provenance of the observational data for our candidate standard stars. The comparison with models, reconciliation with reddening, and the consequent derivation of the full spectral energy density distributions for each of them is reserved for a subsequent paper.

\end{abstract}

\keywords{Stars - Photometric calibration - Standards}


\section{Introduction}\label{sec:intro}
Astrophysics is at the threshold of an era of deep imaging surveys of large portions of the sky, both from the ground and from telescopes in space. 
Projects like the Sloan Digital Sky Survey (SDSS), Pan-STARRS (PS) the Dark Energy Survey (DES), Skymapper, 
ATLAS, ASAS-SN, GALEX and WISE are either complete or in their advanced stages, while the Zwicky Transient Facility (ZTF) has just begun 
and the Large Synoptic Survey Telescope (LSST) is only a few years away.\footnote{A table with a list of all acronyms used in the manuscript is 
presented in the Appendix.}
 \emph{GAIA} and \emph{Kepler} continue to report photometry of millions of stars to very high internal accuracy. These projects have their own native pass-bands and photometric systems with some being 
similar and others differing significantly. 
In order to make the information across these projects commensurate with one another, we must put them on a common 
photometric system. We therefore need a set of calibration references, with which both existing and future surveys 
can be cross-calibrated. Specifically, we seek to establish a set of standard stars which satisfy the following criteria: 

\begin{enumerate}
\item They must have relative spectral energy distributions (SEDs) established to sub-percent accuracy, and preferably to better than half-percent;
\item They must fall within the dynamic range of most, if not all, extant and future deep surveys. 
We surmise that these stars should be fainter than $r \sim$ 16.5 mag, which also puts them within the dynamic 
range of large large-aperture telescopes;
\item They must be distributed across the sky so that they are naturally observed in past, present and future surveys, making it possible 
to retroactively re-calibrate photometry onto a common 
(spectro)-photometric scale. This will allow the direct collation of photometry from different surveys with their own 
respective native pass-bands onto a commensurate platform. For other investigations, a few of the standards will always 
be available from any observatory at any point in time.  
\end{enumerate}

Sub-percent global photometric standardization has proven challenging in the past, but is in high demand for several ongoing scientific endeavors. It is the major source of uncertainty in the use of Type Ia supernovae as probes of the history of cosmic expansion to infer the properties of the dark energy \citep{Betoule2014,scolnic2015,stubbs2015}. Experiments that require accurate and reliable photo-redshift determination, such as weak lensing tomography and baryonic acoustic oscillation analysis with LSST \citep{Gorecki2014}, are also limited by systematic uncertainties arising from their 
relative photometric calibration. 

The chief obstacle for calibrating standard stars with high accuracy from the ground by comparing them to laboratory sources is 
the uncertainty in atmospheric extinction. 
Ground-based survey accuracy is limited by the transmissivity of the atmosphere 
with both chromatic (Rayleigh scattering, ozone, Mie scattering, molecular absorption, aerosol) as well as "gray" (clouds) terms 
varying on small angular and temporal scales. A variety of methods are used to track and account for such effects including 
monitoring (e.g., LIDAR, GPS), and there are many efforts to model the atmosphere (e.g., with MODTRAN, \citealt{burke2014}). 
It would be ideal to place laboratory sources above the terrestrial atmosphere, but this is unlikely to happen in the near future. 
For a more detailed discussion about problems related to obtaining a sub-percent accurate calibration please see 
\citet[hereafter NA16]{narayan2016}.

We therefore seek extra-terrestrial sources for which we can predict the SED to higher accuracy than the uncertainty in predicting 
the transmissivity of the terrestrial atmosphere. 
The best such class of celestial objects we can hope to characterize and model are 
hot DA white dwarfs (DAWDs). These stars have almost pure hydrogen atmospheres, so they are the simplest stellar atmospheres to model: their opacities are known from first principles, at temperatures greater than $\sim$ 20,000K the photospheres are purely radiative, and they are photometrically stable.

The intrinsic DAWD SED can be described by two parameters: effective temperature, $T_{eff}$, and surface gravity, $log(g)$, 
both of 
which can be measured spectroscopically from a detailed analysis of the Balmer line profiles, without using photometry. 
The SED can then be modeled from the ultra-violet (UV) to the near-infrared (NIR) and projected through the transmission 
of any imager or spectrometer at arbitrary resolution. Only the extinction towards the observed 
DAWDs and the overall flux normalization needs to be established.


\begin{figure*}
\begin{center}
\includegraphics[width=0.9\textwidth]{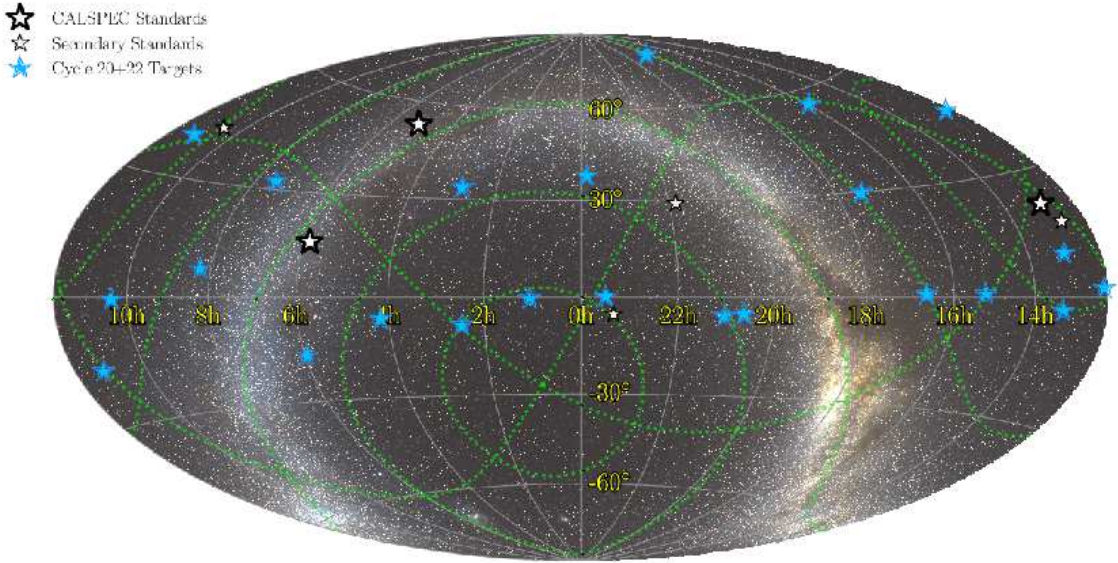}
\caption{A Hammer-Aitoff projection of the 23 candidate spectrophotometric standard DAWDs observed 
with HST in cycles 20 and 22 in equatorial coordinates (blue stars). 
The three \emph{HST} primary CALSPEC standards are marked with a large white star and four spectroscopic flux standards (secondary standards) 
are marked with a small white star. \label{fig:allsky}}
\end{center}
\end{figure*}

\citet{bohlin2000} and \citet[hereafter B14]{bohlin2014a} implemented the pure-hydrogen-WD 
method of flux calibration using three DAWDs, G191B2B, GD153 and GD71 (Hubble Space Telescope (HST) primary WDs).
These stars are brighter than $V \sim$ 13.5 mag, span a range of temperatures 30,000 $\lesssim T_{eff} \lesssim$ 60,000, and are 
un-reddened as a result of their proximity to us. 
B14 found their relative flux distributions to be internally consistent with model 
predictions \citep{gianninas2011, rauch2013} from spectroscopic $T_{eff}$ and $\log g$ to better than 1\% in the wavelength range 0.2 - 0.9
 $\mu$m. 
Spectrophotometry of Vega with STIS \citep{bohlin2004, bohlin2007} referred to the DAWD flux scale shows agreement 
with the \citet{hayes1985} calibration at the 1 to 2\% level, and with the Kurucz\footnote{http://kurucz.harvard.edu/stars/vega/} Vega 
atmosphere model to better than 1\% in the wavelength range 0.5 - 0.8 $\mu$m, but disagrees by 5\% at 0.4 $\mu$m, 
and by 10\% between 0.9 and 1.0 $\mu$m, illustrating the limitation of empirical ground-based methods. 

The internal consistency of the DAWD observations and models ($\leq 5$ mmag) in
the wavelength range 0.2 to 0.9 $\mu$m is superior to the $\sim 2$\% comparison with the best model for 
Vega (Kurucz at $T_{eff} =$ 9,400 K), which is a pole-on rapid rotator with an equatorial dust disk. 
The zero-point for the HST photometric system is defined by the flux of 
3.44$\times$10$^{-9}$ erg$\cdot$cm$^{-2}$$\cdot$ s$^{-1}$ $\cdot$ \AA$^{-1}$ 
for Vega at 0.5556 $\mu$m, as reconciled with the MSX mid-IR absolute flux measures (B14, and \citealt{bohlin2014b}). 
Absolute fluxes for the three \emph{HST} primary WDs are determined by the normalization of their 
modeled SEDs by their respective relative responses to Vega, using STIS precision 
spectrophotometry of all four stars, Vega, G191B2B, GD153 and GD71, and the 
3.44$\times$10$^{-9}$ erg$\cdot$cm$^{-2}$$\cdot$ s$^{-1}$ $\cdot$ \AA$^{-1}$ 
flux of Vega at 0.5556 $\mu$m. This method provides the basis for \emph{HST}'s entire 
calibration system (CALSPEC\footnote{http://www.stsci.edu/hst/observatory/crds/calspec.html}).

\citet{holberg2006} used synthetic photometry of DAWDs in the magnitude range 10 $\lesssim V \lesssim$ 16.5 to place $UBVRI$, 2MASS $JHK$, SDSS $ugriz$ and Str\"omgren $ubvy$ magnitudes on the \emph{HST} photometric scale to 1\%. Later, \citet{holberg2008} confirmed this calibration by using a set of DAWDs in the same magnitude range with well-measured trigonometric parallaxes that agreed at the 1\% level with 
their photometric parallaxes from the Bergeron photometric grid. However, the DAWDs in use to date are still 
too bright for modern deep surveys and large telescopes.

In order to provide flux standards in the dynamic range of large aperture (d $>$ 4m) telescopes, we obtained Wide Field 
Camera 3 (WFC3) \emph{HST} imaging and ground-based spectroscopy for the three \emph{HST} primary (CALSPEC) 
standards, G191B2B, GD153, and GD71, along with 23 DAWDs fainter than $r \sim$ 16.5 mag, at equatorial and northern latitudes.
The need for practical faint standards, 
useful over the optical and near-UV, makes consideration of the effects of interstellar extinction unavoidable. 
Indeed, 
interstellar medium extinction must be incorporated into the definition of the SEDs of all faint flux standards. 
Fortunately, as sub-luminous stars, DAWDs are the optimal choice; simultaneously offering minimal extinction columns and 
wide wavelength coverage, from the far UV to the IR.   


The current paper presents our analysis of photometric and spectroscopic data collected
for the candidate spectrophotometric DAWDs. Preliminary results of the temporal photometric monitoring campaign of the DAWDs are also presented. 
Photometric and spectroscopic data are examined to determine the suitability of each of the 23 candidates as SED standards. 
The joint analysis of photometry and spectroscopy and the derivation of SEDs and reddening to each of these objects is reserved 
for a subsequent paper (Narayan et al.\ 2019, submitted, hereafter NA19).

The structure of the current paper is as follows. In \S 2 we discuss 
the criteria used to select candidate spectrophotometric standard DAWDs and in 
\S 3 we illustrate the photometric observations and the image processing strategy.
In \S 4 we describe the photometric reduction procedures and in \S 5 
the stability monitoring observations for the candidate standards. 
In \S 6 the spectroscopic data reduction strategy is described and in \S 7 we discuss
how our photometry is calibrated and normalized. We summarize the results in \S 8.

\section{Candidate spectrophotometric standard star selection}\label{sec:sele}
Candidate spectrophotometric standard DAWDs were selected from the SDSS \citep{Adelman-McCarthy2008, girven2012, kleinman2013} 
and the Villanova catalog \citep{mccook1999}, with the requirement of being spectral type DA, 
hotter than $\approx$ 20,000 K, and fainter than r $\sim$16.5 mag. We selected an adequate number of stars to 
uniformly cover the sky around the celestial equator and in the Northern hemisphere.
The final sample consists of 23 candidate standard DAWDs.
Table~\ref{table:1} lists the properties of the selected stars and the three \emph{HST} primary CALSPEC standards 
(GD71, GD153, G191B2B), including spectral type, proper motions and distances from the \emph{GAIA} data release 2 (DR2, \citealt{brown2018}).

Fig.~\ref{fig:allsky} shows a Hammer-Aitoff projection of the sky with the distribution of the \emph{HST} primary (CALSPEC) 
standard WDs (large white stars) and the 23 selected candidate standard DAWDs (blue); 
four secondary flux standard stars used to calibrate spectra analyzed in the current paper 
are also shown in the figure as small white stars. The figure shows that candidate standards have an homogeneous 
coverage over the Northern hemisphere and the celestial equator with approximately 1 star every 2 hours. 

We have a sample of candidate DAWDs for the Southern hemisphere for which spectra collected with the 
Goodman spectrograph on the SOAR telescope (CTIO) are available and \emph{HST} photometry was 
collected during Cycle 25 (GO-15113, PI: Saha). 
A subsequent paper will present photometry and spectroscopy for these new candidates.
The final goal is to provide an all-sky set of sub-percent precision  
spectrophotometric standards so that at least 3 of these stars are visible at any time from any observatory at airmass less than 2.

\begin{figure}
\begin{center}
\includegraphics[height=0.3\textheight,width=0.5\textwidth]{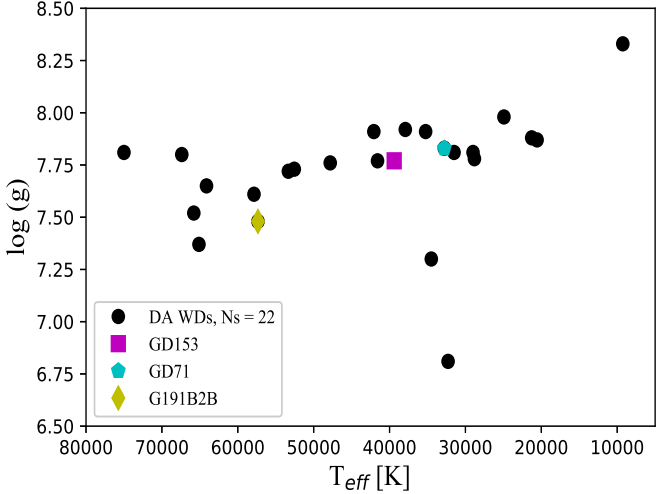}   
\caption{$T_{eff}$ vs $log(g)$ plane for 22 selected DA white dwarfs (black dots) and the three \emph{HST} primary CALSPEC 
standards G191B2B (yellow diamond), GD153 (magenta square), and GD71 (cyan pentagon). \label{fig:logg}}
\end{center}
\end{figure}

Preliminary effective temperatures and gravities to select candidate spectrophotometric standards 
were retrieved from the SDSS and the Villanova catalogs. 
The ground-based spectra collected by us using different facilities will be used to derive 
more accurate temperatures and gravities for all the DAWDs as described in section~\ref{sec:spectra}.
The \emph{HST} primary WDs and the candidate standards are plotted in the $T_{eff}$ vs $log(g)$ plane in Fig.~\ref{fig:logg}. 
For star WD0554-165, $T_{eff}$ and $log(g)$ measurements are not available in the literature.
The figure shows that star SDSS-J172135.97+294016.0 is much cooler compared to the
other DAWDs, with an effective temperature of $T_{eff}$ = 9,261 K(see Table~\ref{table:1}). 
This star was included in the sample because of an early decision before we restricted ourselves to purely
radiative atmospheres with temperatures $T_{eff} >$ 20,000 K. The observations and data reduction for this star 
were carried through, but this object will no 
longer be included in the network of standard DAWDs.

We matched the list of candidates with PS Data Release 2 
catalog \citep{Chambers2016, Flewelling2016} and obtained $g,r,i,z$ aperture photometry 
magnitudes for all the stars. These data will be later used to check the the stellar density around the 
candidate standards.
We also retrieved \emph{GAIA} DR2 $G, Bp, Rp$ magnitudes for the DAWDs. 
PS and GAIA sets of magnitudes are listed in Tables~\ref{table:2} and \ref{table:3}, respectively.

\begin{figure}
\begin{center}
\includegraphics[height=0.3\textheight,width=0.5\textwidth]{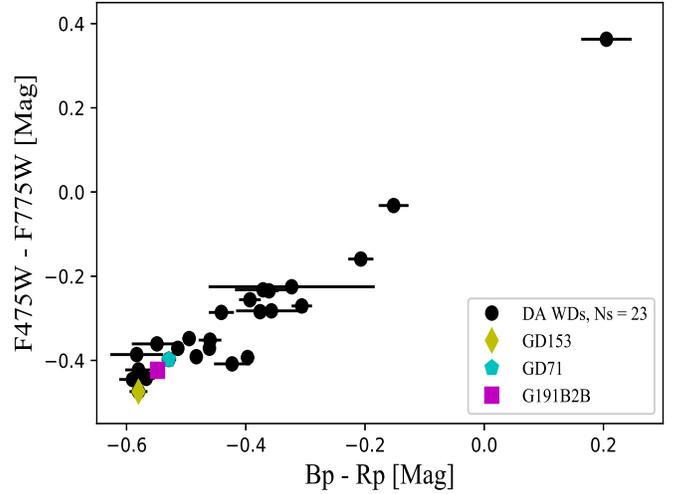}  
\caption{$Bp - Rp,\ F475W - F775W$ color-color diagram for the 23 selected DA white dwarfs (black dots) and the three \emph{HST} 
primary standards, G191B2B (yellow diamond), GD153 (magenta square), and GD71 (cyan pentagon).\label{fig:dao_gaia}}
\end{center}
\end{figure}

Fig.~\ref{fig:dao_gaia} shows the $Bp - Rp, \ F475W - F775W$ color-color diagram for the 23 
candidate standard DAWDs (black dots) and the three primary standards, G191B2B (yellow diamond), 
GD153 (magenta square) and GD71 (cyan penthagon). 
Photometry in the \emph{HST} filters is from this work (see section~\ref{sec:photo} and Table~\ref{table:9}).
The selected candidate standards and the primary  \emph{HST} WDs cluster along a well-defined sequence and have 
colors in the range -0.6 $\lesssim Bp - Rp \lesssim$ -0.1 and -0.4 $\lesssim F475W - F775W\lesssim$ 0 mag,
excluding star SDSSJ172135.97+294016.0, $\sim$ 0.3-0.4 mag redder than the rest of the WDs, which
is the cool DAWD included in our sample, as explained before.



 \section{Photometric observations}\label{sec:obs}
Photometric data discussed in this investigation were collected with the WFC3 UVIS and IR cameras 
on board the \emph{HST} during Cycle 20 and 22 (proposals GO-12967 and GO-13711, PI: Saha).
Observations were taken in five filters in Cycle 20, namely $F336W$, $F475W$, $F625W$,
$F775W$, $F160W$. In Cycle 22 the near UV filter $F275W$ was added, to better characterize the line-of sight extinction and the reddening law towards the observed stars.

Nine of the candidate DAWDs are distributed along the celestial equator (hereinafter {\it equatorial} DAWDs) and were observed in both Cycle 20 and Cycle 22, while the other fourteen DAWDs and the three \emph{HST} primary CALSPEC standards were observed only in Cycle 22.
Cycle 20 observations of the nine {\it equatorial} DAWDs were used in NA16 to demonstrate the effectiveness of our method 
to establish a network of spectrophotometric standards. We repeated the observations of these stars in Cycle 22 to improve the 
precision of the photometry. Moreover, photometry of the {\it equatorial} DAWDs will be used
to determine the photometric offset between the two observing cycles. 
The \emph{HST} primary CALSPEC WDs were observed in Cycle 22 to allow us 
to directly tie the photometry of the 23 DAWDs to the HST photometric system.

Exposure times for our observations range from 1 to 220s for the $F275W$ filter, 
1 to 160s for $F336W$, 1 to 160s for $F475W$, 1 to 355s for $F625W$, 1 to 680s for $F775W$, and 3 to 499s for $F160W$. 
Table~\ref{table:4} lists the log of the observations for Cycle 20 and 22.

Observations span a time interval of about 1 year for Cycle 20 (November 
2012 until September 2013) and about 1.3 year for Cycle 22 (September 2014 until January 2016),
with the \emph{HST}  primary CALSPEC WDs observed at the beginning and the end of Cycle 22 to track 
the change in sensitivity of the telescope and instrument system.

\begin{figure*}
\begin{center}
\includegraphics[height=0.40\textheight,width=0.75\textwidth]{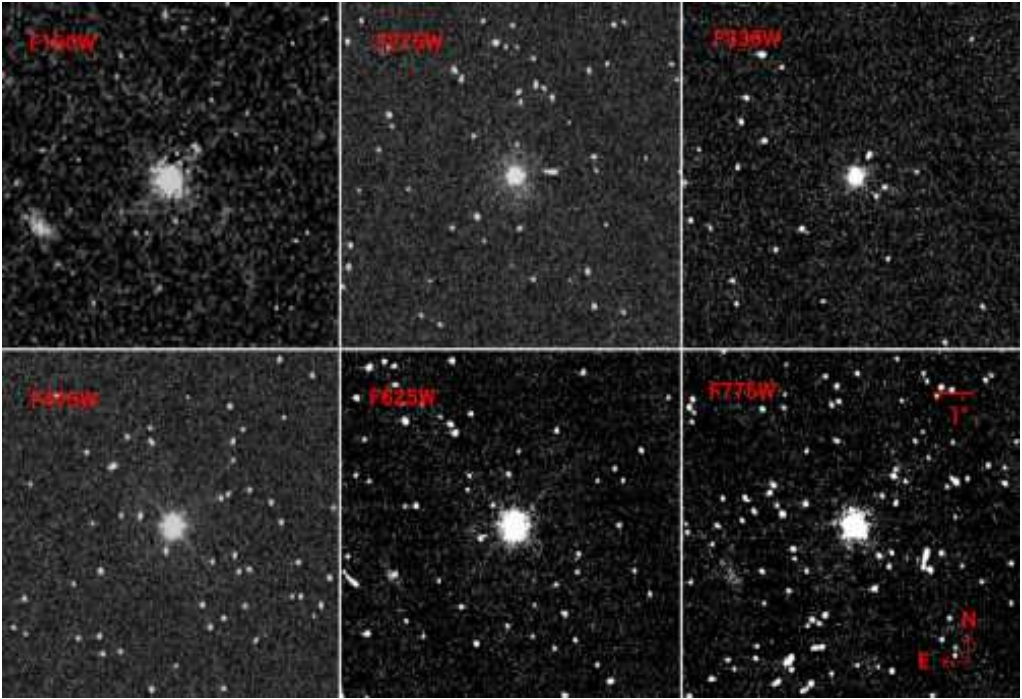} 
\caption{WFC3 FLC images in six filters for the DA white dwarf SDSSJ232941.330+001107.755. 
The scale of the image and the North and the East directions are shown in the bottom right panel.
CRs and different detector artifacts, such as hot and bad pixels are visible in all images. \label{fig:wd2327-000}}
\end{center}
\end{figure*}

For the nine {\it equatorial} DAWDs, three dithered exposures in $F336W$ and $F475W$, 
and two exposures for the other filters were collected in Cycle 20. 
The same targets were observed in Cycle 22 with a cosmic ray split of 3 exposures for 
$F275W$, and one exposure was added for the other filters.
Summarizing, a total of three exposures for each of the WFC3 filters were collected for the nine {\it equatorial} DAWDs.
At least three exposures per filter are needed to check consistency in the photometry at sub-percent precision: WFC3 images are  
affected by cosmic rays (CRs) and different detector artifacts, such as hot and bad pixels, blobs, ghosts. 
As an example, images in six filters for star SDSSJ232941.330+001107.755 are shown in Fig.~\ref{fig:wd2327-000}.
If a star has discrepant measurements in two exposures, the third image will allow us
to identify the outlier measurement and discard the affected exposure. 

The other fourteen WDs were observed only in Cycle 22, by collecting 5 to 7 dithered exposures for each filter. 
The \emph{HST}  primary CALSPEC standards were observed in Cycle 22 with 6 to 8 exposures 
per visit, for a total of 3 visits (18 to 24 exposures) in all filters, spanning an average time interval of about 1 year and 3 months, 
from September / November 2014 to November 2015 / January 2016.
For more details on the observation strategy please see the log in Table~\ref{table:4}.

Parallel observations with the Advanced Camera for Surveys (ACS) in the $F475W$ and $F775W$ filters 
were collected, including stars a few arcminutes away from the candidate spectrophotometric standards.
The analysis of these images and an evaluation of the usefulness of the 
observed stars as supplementary standard stars will be presented in a forthcoming paper.

\subsection{Image processing}\label{sec:data}
WFC3 UVIS images for Cycle 20 and Cycle 22 were processed with version 3.3 of the WFC3 calibration pipeline 
({\it cal\_wf3}) that treats the two chips, UVIS1 and UVIS2, individually \citep{deustua2017}. 
The image photometry reference table (IMPHTTAB) used is {\it z7n21066i\_imp.fits} and writes PHOTFLAM 
values for a 10 pixel aperture \citep{deustua2016}. A newer IMPHTTAB file was released in June 2017, which 
provides PHOTFLAM values for an infinite aperture, but it was not used to reduce our dataset.
Cycle 20 images were collected by using the full UVIS1 aperture (UVIS1-FIX), with the target star placed in its center. 
Few pixel dithered exposures were collected to correct for detector artifacts and CRs 
(see the observation log in Table~\ref{table:4}). The sub-array UVIS2-C512C-SUB aperture was used for Cycle 22 
observations, with the target star placed in its center and the exposures dithered by a few pixels (less than $\sim$ 20). 
This allowed us to place the WDs closer to the read-out amplifier to mitigate the charge transfer inefficiency effects.

Starting from version 3.3, the WFC3 calibration pipeline scales UVIS2 fluxes to the UVIS1 chip by default. Therefore, 
we manually re-processed all the images with {\it cal\_wf3} by setting {\it FLUXCORR = OMIT} in the header 
to avoid the flux scaling and to keep the photometry on the UVIS2 detector system. 
The scale factor between Cycle 20 UVIS1 and Cycle 22 UVIS2 photometry will 
be estimated and applied to the measured magnitudes later (see Section~\ref{sec:offset}).

All WFC3-UVIS images were corrected for Charge Transfer Efficiency (CTE) by using the 
official WFC3 software\footnote{$http://www.stsci.edu/hst/wfc3/ins\_performance/CTE$} and 
the WFC3 Pixel Area Map (PAM) was 
applied\footnote{$http://www.stsci.edu/hst/wfc3/pam/pixel\_area\_maps$} to correct for differences in the area of each 
pixel in the sky due to the geometric distortion of WFC3-UVIS.

WFC3 infrared (WFC3-IR) images were collected by using the full camera aperture (IR-FIX) and placing the 
targets at the center of the detector with every exposure dithered by 10 to 20 pixels. 
This strategy was used to avoid self-persistence.
For the three primary WDs, observations were collected by using 
the IRSUB256-FIX and the IRSUB512-FIX sub-apertures to allow more exposures in the same orbit.
Images were processed with the {\it cal\_wf3} calibration pipeline and the WFC3 PAM was applied.

\section{Optimal extraction of the white dwarf photometry}\label{sec:photo}
Following a series of tests, we have discovered that the individual WFC3 images cannot be combined in the standard 
pipeline reduction and be expected to yield measurements with milli-mag level uncertainties. 
In particular, the anti-coincidence method of eliminating CR events can affect the cores of stars. 
In reducing Cycle 20 WFC3 data, NA16 found that combining images with the drizzle algorithms (Drizzle Pac\footnote{$http://drizzlepac.stsci.edu/$}), 
by using the pipeline default input parameters, is not suitable for our purpose. While drizzling eliminates CRs, 
it introduces noise by over-correcting for differences in the cores of bright stars. 
Therefore, in order not to compromise the quality of the good images, where the target star is unaffected by a 
CR in its measuring aperture, we had to manually discard 
any image with a CR event over the measurement aperture on those few occasions it happened (3-4 \% of the total number of images).

Our current reduction strategy thus involved performing photometry on the individual CTE-corrected images (FLC) 
for the WFC3-UVIS detector, and on the FLT images for WFC3-IR, after having applied the PAM correction.
We had to check all the $\sim$ 800 images for the presence of a CR event inside the star aperture radius: this was a rather 
tedious procedure but it allowed us to obtain a sub-percent accurate photometry for our set of standard stars.

\begin{figure*}
\begin{center}
\includegraphics[height=0.9\textheight,width=0.9\textwidth]{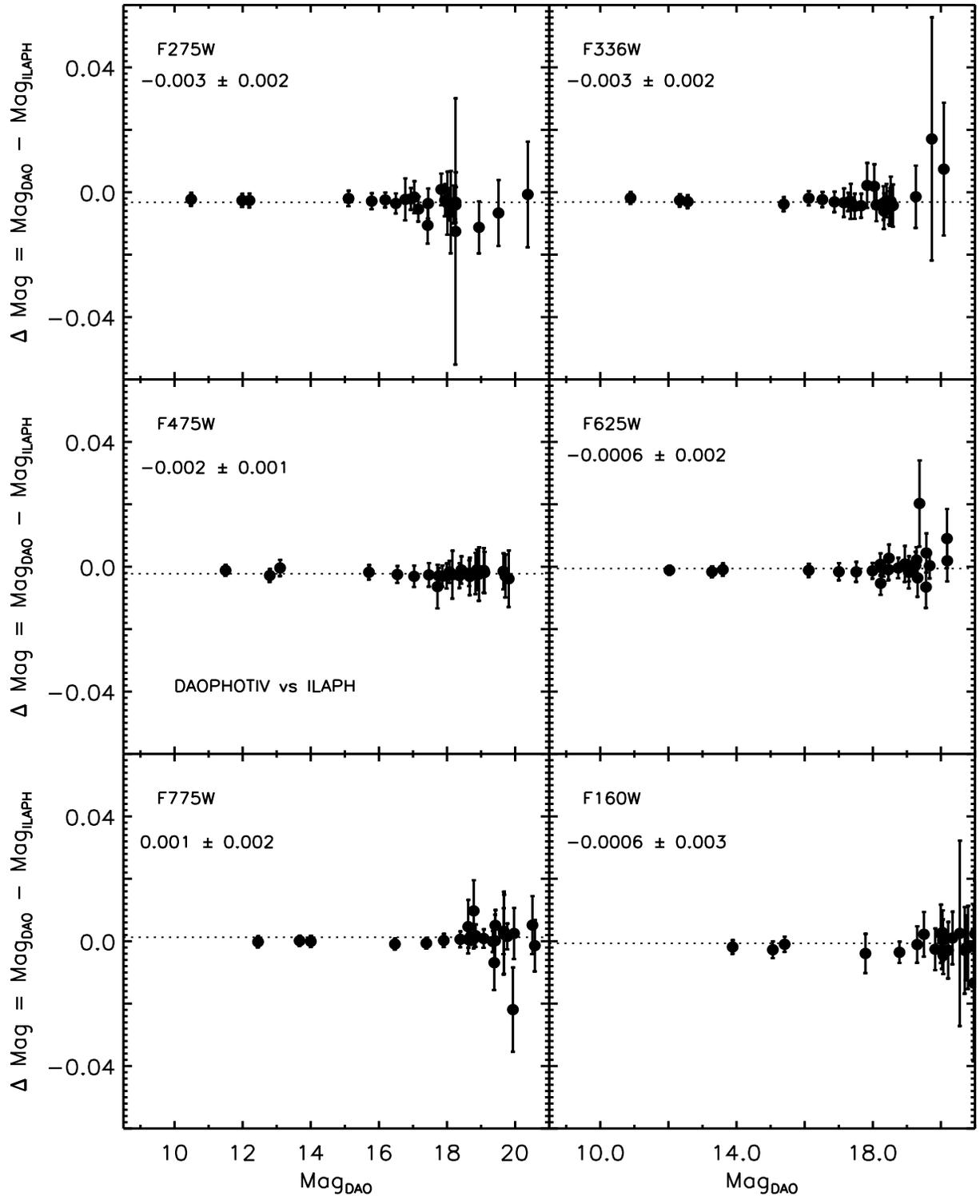}    
\caption{Comparison of DAOPHOT and ILAPH aperture photometry magnitudes, $\Delta$ Mag = Mag$_{DAOPHOT}$ - Mag$_{ILAPH}$, 
in six filters as a function of DAOPHOT magnitude, for the 23 candidate spectrophotometric DAWDs and the 3 \emph{HST} primary CALSPEC standards. The mean magnitude differences with the relative dispersions are shown. \label{fig:delta_ilaph}}
\end{center}
\end{figure*}

We used three completely independent packages to perform aperture photometry on all the images: 

1) Source Extractor \citep[hereafter SExtractor]{bertin1996};
2) DAOPHOTIV \citep[hereafter DAOPHOT]{stetson1987};
3) ILAPH (an IDL based interactive program for aperture photometry and growth curve analysis written by A.~Saha, 
customized for the data at hand).

Performing photometry by using different packages may generate systematic differences in the results.
Our strategy will then enable us to track down possible systematic issues due to the data reduction method used. 
Moreover, it will allow us to determine realistic uncertainties, which is of paramount importance for our study,
 since analysis downstream depends on these for weights. 

After analyzing photometric growth curves for a sample of faint and bright WDs, we found the optimal aperture radius 
for the photometry to be 7.5 pixels for the WFC3-UVIS images and 5 pixels for the WFC3-IR images, i.e. $\sim$ 0\farcs3~ and $\sim$ 0\farcs65, respectively.
The local sky background for each source was estimated in a rectangular region (SExtractor) and a circular (DAOPHOT, ILAPH) annulus around the aperture centered on the star. In the case of SExtractor the box has a size of 20 pixels and the sky background in this region is estimated as a modified mode (2.5$\times$Median - 1.5$\times$Mean) after an iterative sigma-clipping rejection of the outlier pixels. DAOPHOT uses the mode (3.0$\times$Median - 2.0$\times$Mean) as the best sky estimator after an iterative sigma-clipping rejection of the outliers. However, in non-crowded stellar fields, 
if the mean sky value in the selected annulus is smaller than the median, than the mean of the sky value is used as best sky estimate. In this case, we used an annulus with radii $r_{in}$ = 156 and $r_{out}$ = 165 pixels around the target DAWDs. We selected these values since 156 pixels corresponds to $\sim$ 6\arcsec~ and it can be considered to infinity relative to the star's position.  

ILAPH was configured to use the median sky value in the selected annulus around the stars as the best sky estimator. 
This is more robust than the mean, since the latter is vulnerable to the presence of contaminating objects in the annular sky aperture.
For UVIS, an aperture radius of 7.5 pixels and an annulus with radii $r_{in}$ = 20 and $r_{out}$ = 30 pixels was used for the sky. 
While this measurement procedure disregards light outside the respective apertures due to the extended {\it skirt} of 
the stellar Point-Spread-Function (PSF), it is asserted that the {\it skirt} affects all stars equally, and as long as we measure 
the bright \emph{HST} primary CALSPEC standards in exactly the same way as our target DAWDs, we are measuring instrumental 
magnitudes that all share a common zero-point offset. 
For the IR images ($F160W$), ILAPH was used in the same way, but with aperture radius of 5 pixels and a 
sky aperture annulus from 14 to 21 pixels. 
The sky apertures were chosen with some experimentation: stability in the measured instrumental magnitude 
values from image to image for the same object was used as the criterion for selection. 
The program also looks at the pixel to pixel scatter within the annular sky aperture, and propagates that variance
into the measurement error estimate. 
ILAPH was customized to use the actual fluctuation in the sky background, not from just the shot noise 
(Poisson statistics) of the adopted sky brightness, in the calculation of photometric uncertainty. 
Subsequent analysis utilizes the uncertainties as weights, so it is crucial to get this estimate to be as realistic as possible.

An accurate estimate of the sky background is fundamental for our analysis. In particular, a wrong estimate of 
the sky background has a greater effect on the faintest DAWDs and can introduce a systematic bias in the measurements. 
Fig.~\ref{fig:delta_ilaph} shows the comparison between the measurements in the six filters, 
$F275W$, $F336W$, $F475W$, $F625W$, $F775W$, and $F160W$, obtained with DAOPHOT and 
ILAPH ($\Delta$ Mag = Mag$_{DAOPHOT}$ - Mag$_{ILAPH}$) for all the observed DAWDs as a function of the 
measured DAOPHOT magnitude. The single epoch magnitudes, corrected for the instrumental effects, 
including the sensitivity difference between Cycle 20 and Cycle 22, were averaged as described in Section~\ref{sec:photo}.

\begin{figure*}
\begin{center}
\includegraphics[height=0.9\textheight,width=0.9\textwidth]{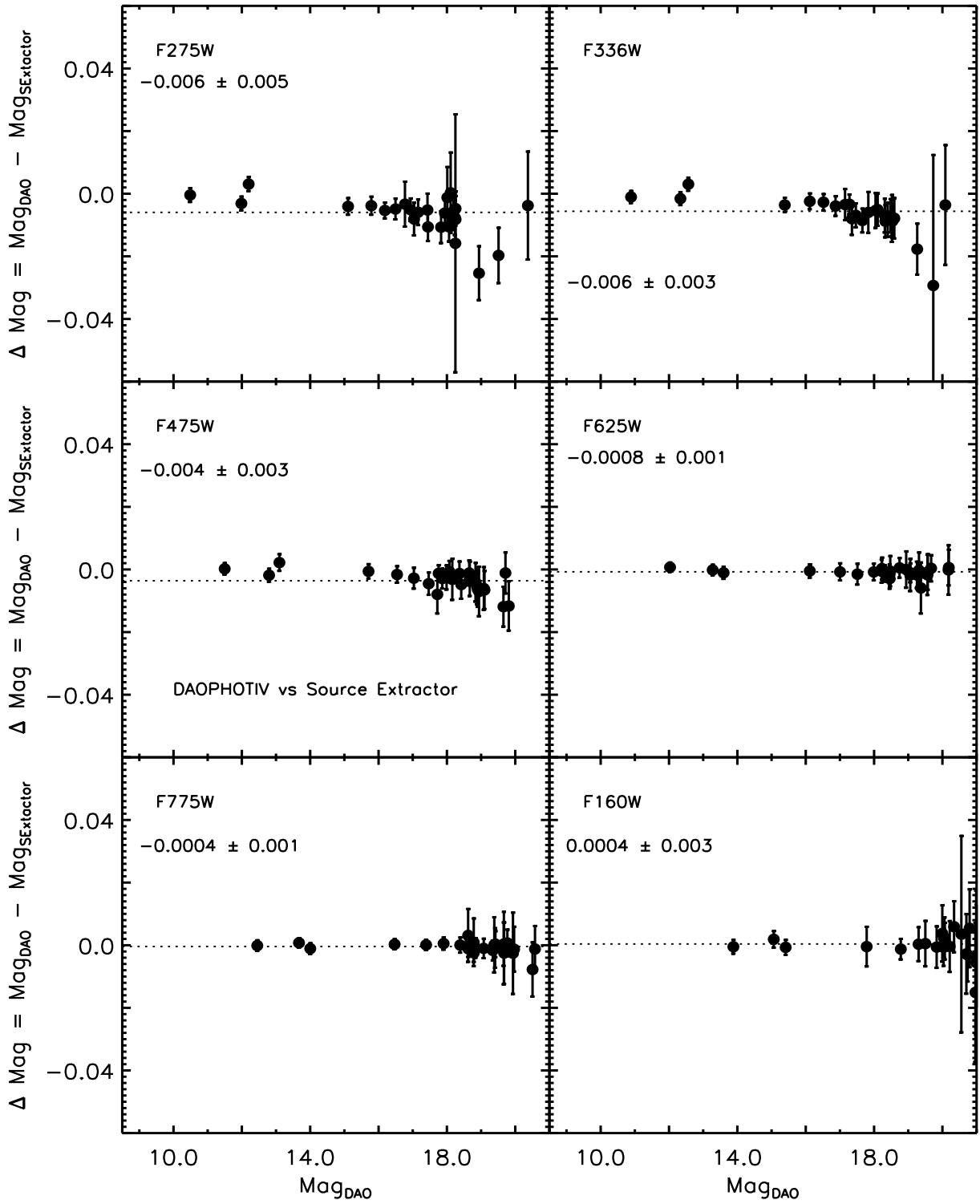}  
\caption{Same as Fig.~\ref{fig:delta_ilaph} but for DAOPHOT and SExtractor magnitudes, $\Delta$ Mag = Mag$_{DAOPHOT}$ - Mag$_{SExtractor}$. \label{fig:delta_sex}}
\end{center}
\end{figure*}

The plot shows that measured aperture magnitudes with DAOPHOT and ILAPH are, within uncertainties, 
in very good agreement, with a biweight mean difference less than 0.003 mag in all filters, and with a dispersion 
less than 0.003 mag for all UVIS filters and less than $\sim$ 0.003 mag for $F160W$.
A couple of stars, SDSSJ102430.93-003207.0 and SDSSJ172135.97+294016.0, show very large uncertainties, 
$\sim$ 0.02 and 0.04 mag, respectively, in the $F275W$ and the $F336W$ filters.
The first DAWD was problematic already in NA16 Cycle 20 measurements, and is a candidate variable (see Section \ref{sec:time}).
Star SDSSJ172135.97+294016.0. has no clear problems on the images, but an undetected faint CR falling on the aperture radius could be the culprit.
However, this star is already excluded from our set of spectrophotometric standards due to its low effective temperature ($T_{eff} \sim$ 9,000 K),
so we do not investigate this issue further.

The same comparison is performed for all the measurements obtained with DAOPHOT and SExtractor in Fig.~\ref{fig:delta_sex}, where $\Delta$ Mag = Mag$_{DAOPHOT}$ - Mag$_{SExtractor}$. In this case, the measurements agree quite well within uncertainties but on average, they have a larger dispersion, up to 0.005 mag for $F275W$. 
Moreover, a slight trend with magnitude is present, with SExtractor magnitudes being fainter at fainter magnitudes.

\begin{figure*}
\hspace{-1.5cm}
\includegraphics[height=19.5cm,width=21.5cm]{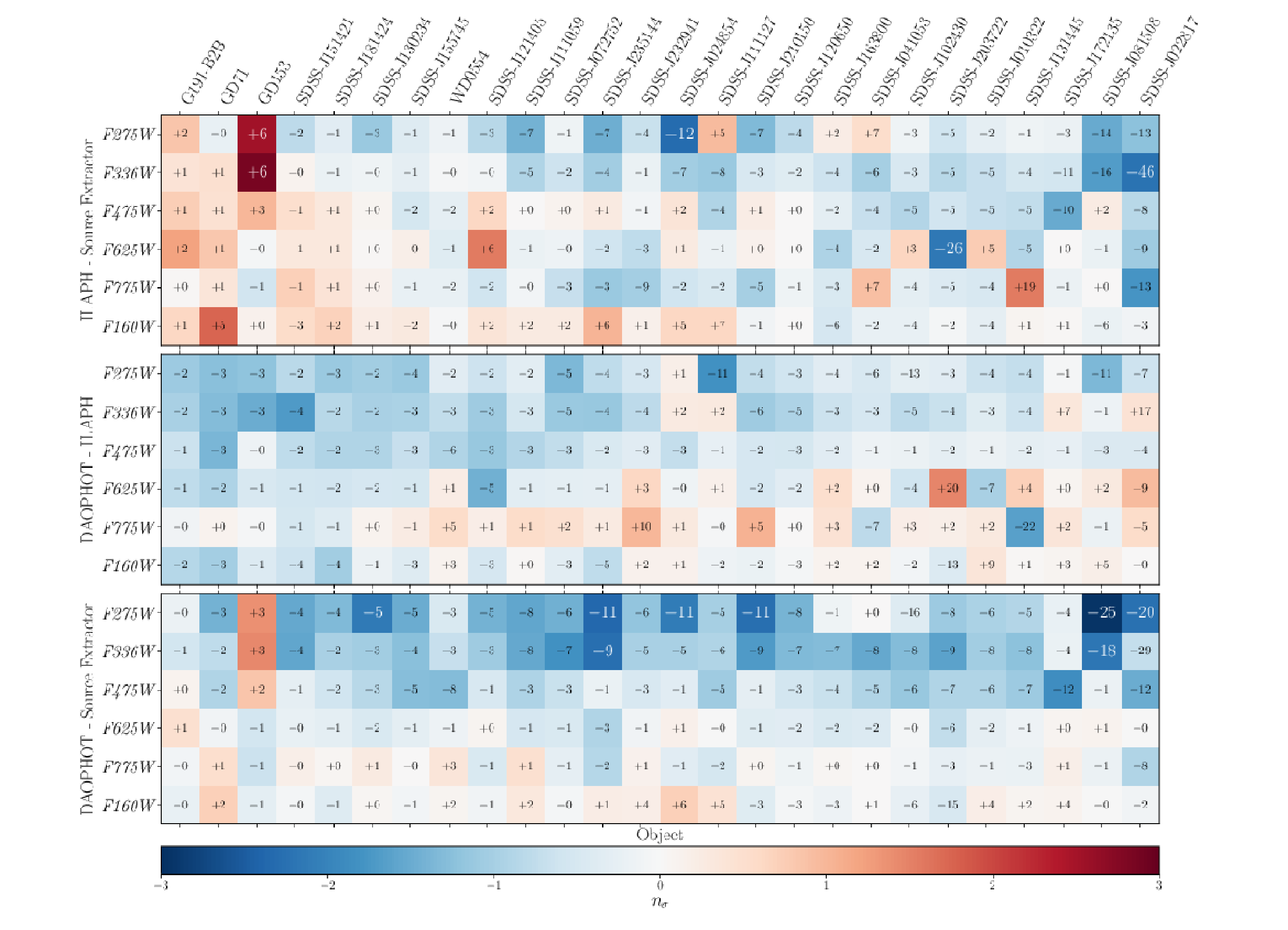}
\caption{\label{fig:matrix} Comparison matrix between magnitudes measured with ILAPH, DAOPHOT and Source Extractor for 
the 23 candidate standard DAWDs and the 3 \emph{HST} primary CALSPEC stars in all the WFC3 observed filters. 
The color scale is labeled at the bottom of each panel and the name of the stars is marked at the top of the panels. 
Stars are sorted by $g$-band magnitude, starting with the brightest on the left. Filters are labeled on the left of each panel. 
The color of the box is bluer when the weighted magnitude difference between the two methods is negative, i.e., 
when the magnitude of the first labeled method is brighter compared to the magnitude of the second method, 
while is red when the weighted magnitude difference is positive.
See text for more details.}
\end{figure*}

To further investigate this issue, we produced a matrix comparison of the three data reduction methods in Fig.~\ref{fig:matrix}. 
Each depicted box is color-scaled according to the weighted magnitude difference between two of the three methods, estimated as:

\begin{equation}
\Delta Mag = (Mag_{meth1} - Mag_{meth2})/\sqrt{err_{meth1}^{2} + err_{meth2}^{2}}
\end{equation}

where $Mag_{methX}$ and $err_{methX}$ are the magnitudes and magnitude errors for each method, respectively.

Each box corresponds to a star measured in one of the six filters, sorted by magnitude 
(brightest on the left of the matrix), and the numeric text value is the magnitude difference, in milli-mag, between the measurements of the two methods.
The color of the box is bluer when the weighted magnitude difference between the two methods is negative, i.e., when the magnitude of the first labeled method is brighter compared to the magnitude of the second method, while is red when the weighted magnitude difference is positive. The text in the boxes is larger and white when the difference in magnitude between the two methods is larger than 2$\sigma$, i.e., significant compared to photometric errors.

The middle panel of the figure confirms that ILAPH and DAOPHOT provide very comparable results within the uncertainties, 
with no significant magnitude difference for all the measured stars in all filters. However, for fainter stars (right section of the matrix), 
a very slight trend with color is present, in the sense that these stars are fainter in the redder filters ($F625W$, $F775W$, and $F160W$) 
when measured with DAOPHOT (redder boxes on the bottom right of the panel). 
On the other hand, this trend is well within the uncertainties of the measurements.

The top and the bottom panels of Fig.~\ref{fig:matrix} confirm that SExtractor magnitudes are systematically fainter 
compared to ILAPH and DAOPHOT magnitudes in all filters, as suggested by Fig.~\ref{fig:delta_sex} (bluer boxes in the panels).
This difference has either a magnitude and a color effect: 1) magnitudes for brighter stars seems to agree between SExtractor
and the other two methods, or to be brighter when measured with SExtractor (left part of the panels), while 
fainter stars are fainter when measured with SExtractor (right part);
2) the discrepancy is larger for the bluer filters ($F275W$ and $F336W$, top part).

To understand the current discrepancy, we compared the sky background values estimated with the three different methods. 
SExtractor local sky background shows systematically higher values when compared to DAOPHOT and ILAPH sky values. 
This difference in sky values is larger at fainter magnitudes, making SExtractor magnitudes fainter for fainter stars. 
On the other hand, DAOPHOT and ILAPH sky values agree quite well within the uncertainties.

\subsection{IR photometry}
The WFC3-IR detector is affected by persistence, i.e. the residual signal of a large 
incident light level that can last on the images from minutes to days \citep{long2011,long2013,gennaro2018}.
Some of our observations could have been scheduled after IR observations 
that cause persistence, or bright objects in the images could cause 
persistence in the same exposures. To verify this we checked all images 
for the level of external (due to previous observations), or internal (due to objects in the same exposure)
persistence by using the available WFC3 persistence tool\footnote{https://archive.stsci.edu/prepds/persist/search.php}. 
The search revealed that none of our observations is heavily affected by external or
internal persistence, with the fraction of pixels with a residual signal larger 
than 0.01 e$^{-}$/s being less than 0.1\%. Note that the dark current signal for 
WFC3-IR is 0.04 e$^{-}$/s. 
Three visits, one for star G191B2B, one for GD153 and one for 
SDSSJ181424.075+785403.048, have a fraction of pixels with a residual signal larger 
than 0.01 e$^{-}$/s of 0.16, 0.25 and 0.33\%, respectively, due to external persistence.
However, the affected pixels do not overlap with the target DAWDs location, being more than 50 pixel away.
Internal persistence is not an issue for all our observations.
Brighter stars, such as the three primary WDs, could cause self-persistence on the IR images.
In order to avoid that, we dithered each exposure by more than 10 pixels as recommended by
the WFC3 team.

Another issue affecting IR observations is the count-rate non-linearity (CRNL),
that is the non-linearity of the detected counts with the total incident flux 
on the camera. This effect can be relevant for our observations since the target DAWDs
cover a range of more than 10 magnitudes. 
The CRNL was characterized for the WFC3-IR camera
by \citet{riess2010a}, \citet{riess2010b}, and \citet{riess2011}, and who measured 
0.010$\pm$0.0025 mag per dex for the $F160W$ filter. The net effect of CRNL is that 
photometry of very faint stars, i.e. background dominated, appears 
fainter \citep{riess2010b}. An accurate characterization of the CRNL for the program IR photometry, 
based on the observations and on models, will be provided in NA19.
We do not apply any CRNL corrections on the photometry presented in this paper.

\subsection{Shutter shading}
The accuracy of the measured magnitudes on WFC3-UVIS could be affected by 
the shutter shading effect. For the brightest stars in our sample we used very short 
exposures times ($t <$ 2s). For these short times, shutter vibration can affect 
the actual duration of the exposures, leading to fainter measured magnitudes on the image. 
This effect was studied in detail by the WFC3 team and discussed in different Instrument Science Reports \citep{hilbert2009,sabbi2009,sahu2014,sahu2015}. Shutter vibrations in short exposures also
results in a broadening of the observed PSF. When observing by using the shutter blade B, 
the Full-Width-Half-Maximum of the stellar images is systematically larger than when 
using the blade A. The larger shutter vibrations when using blade B 
can introduce a flux measurement uncertainty up to $\sim$ 2\% for photometry performed
with aperture radii smaller than 5 pixels \citep{sabbi2009}.

\begin{figure}
\begin{center}
\includegraphics[height=0.35\textheight,width=0.5\textwidth]{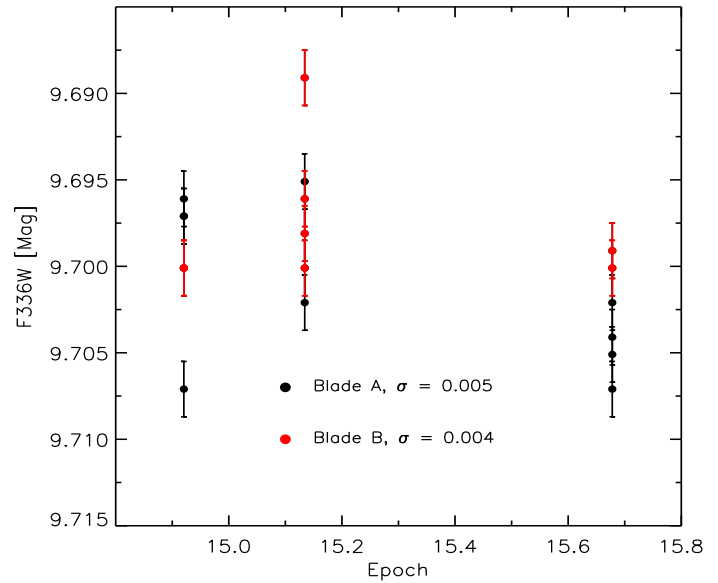}  
\caption{Aperture magnitudes in the $F336W$ filter for star G191B2B 
measured with the shutter blade A (black points) and B (red) plotted versus the epoch.
Error bars are labeled. The standard deviation of the measurements for the two blades is also shown.\label{fig:shutter}}
\end{center}
\end{figure}

For our WFC3-UVIS data, we used an aperture radius of 7.5 pixels and so 
shutter shading should not affect images collected by using blade B. 
However, we checked for the presence of this effect on the images of the brightest of the 
primary CALSPEC standards, G191B2B. The selected $F336W$ images were 
collected in a sequence of 1.0 second exposures alternating the two shutter blades (ABAB...).
Fig.~\ref{fig:shutter} shows instrumental aperture magnitudes in the $F336W$ filter 
for G191B2B measured with the shutter blade A (black points) and B (red) plotted versus the observing
epoch.The standard deviation of the measurements is 0.005 and 0.004 mag for blade A and B, respectively. 
The plot shows that there is not significant difference between magnitudes measured 
on images collected by using blade A or B.
We performed the same experiment for G191B2B images collected with the other 
UVIS filters and obtained similar results. 

The exposures times for the other DAWDs are longer than for G191B2B (see the observing log 
in Table~\ref{table:4}), so we did not verify for shutter blade effects in all the other observations and
we can safely assume that none of the exposures is affected.

\subsection{Testing photometry for crowding effects}
Another factor that could affect the accuracy of measured magnitudes is the presence
of unseen neighbor stars in the DAWD aperture radius. To test this hypothesis, we performed 
some artificial star (AS) tests simulating stars of different brightness with centroids from 0 to 5 pixels 
away from the DAWD. We simulated stars from 3 to 7 magnitudes fainter than the DAWDs. 
Results of the simulations show that AS more than 6 mag fainter than the target DAWD falling in 
the 7.5 pixel aperture radius do not affect the measured magnitudes. 
On the other hand, brighter neighbor stars falling inside the aperture radius
could affect the photometry of the target DAWD by adding $\sim$ 1\% of noise to the measurement.

However, our candidate standard DAWDs are in sparsely populated stellar fields and the observed 
WFC3 sub-array field of view (FoV) is $\sim$ 20\arcsec. We checked the PS catalog to look for 
the presence of other stars in the observed FoV and found only the target DAWDs or a maximum 
of other two objects (well outside the aperture radius) down to the PS detection limit ($g \sim$ 23 mag),
so more than 5 magnitudes fainter than our targets.
We can then safely assume that the photometry of the DAWDs is not affected by contamination of 
unseen neighbor stars.

\subsection{Magnitude offset between the \emph{HST} observing cycles}\label{sec:offset}
Images in the five filters $F336W$, $F475W$, $F625W$, $F775W$, and $F160W$ were collected in Cycle 20 and Cycle 22 for the nine {\it equatorial} WDs. For this subset of targets we then have two sets of measurements. Because the primary CALSPEC WDs, which anchor our photometry to the \emph{HST} system, were only observed in Cycle 22, we need to estimate the magnitude offset between the two cycles and calibrate Cycle 20 measurements to Cycle 22.
Cycle 20 observations were performed by using the full UVIS1 aperture, while
Cycle 22 exposures were collected with a UVIS2 sub-array. The magnitude offset needs to take into account the difference due to observing with two different WFC3 detectors and all the effects due to observations taken more than 2 years apart. 

\begin{figure}
\begin{center}
\includegraphics[height=0.65\textheight,width=0.5\textwidth]{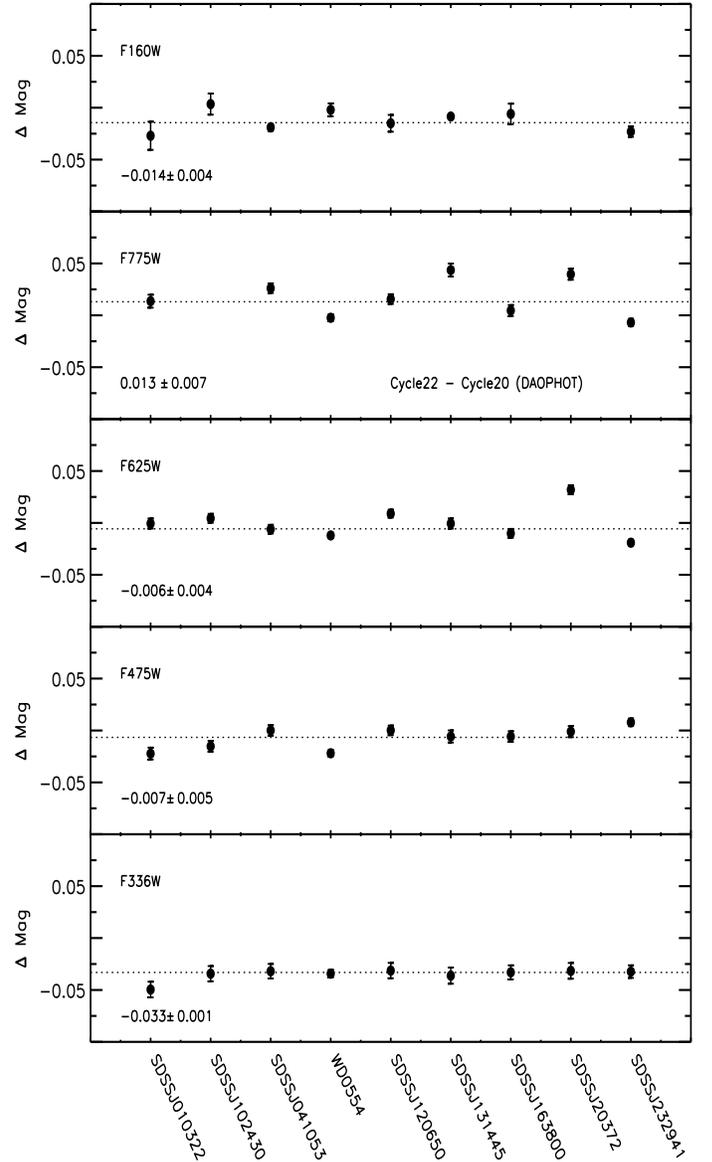}
\vspace{0.25cm}
\caption{Comparison of Cycle~20 and Cycle~22 average magnitudes measured with DAOPHOT in six filters, namely $F275W$, $F336W$, $F475W$, $F625$, $F775W$, and $F160W$ for nine {\it equatorial} WDs.
The weighted mean magnitude differences between the two observing cycles with the relative errors are also labeled. \label{fig:offset}}
\end{center}
\end{figure}

Fig.~\ref{fig:offset} shows the comparison between Cycle 20 and Cycle 22 magnitudes in 
the five filters for the nine {\it equatorial} WDs. Star SDSSJ102430.93-003207.0 has very discrepant
measurements in the $F775W$ filter ($\Delta$ Mag $\sim$ -0.21); 
NA16 claim that this WD might be variable. The same applies to star SDSSJ203722.169-051302.964, 
where the $F160W$ measurements are in strong disagreement between the two observing cycles ($\Delta$ Mag $\sim$ 0.27).
The spectrum of this WD shows emission features in the core of the Balmer absorption lines, indicating the presence of a low-mass companion star (see \S~\ref{sec:spectra} for more details). Both stars were removed from the sample to estimate the magnitude offset.

The offsets are between $\sim$ 0.005 and $\sim$ 0.03 mag, depending on the filter, with an average 
dispersion of 0.005 mag. Column 8 and 9 of Table~\ref{table:6} list the magnitude offsets and their uncertainties 
between the two \emph{HST} observing cycles in the different filters.

After having applied these magnitude offsets, we estimated the weighted mean instrumental magnitudes for 
all the 23 observed DAWDs based on the photometry of the two cycles.

\section{Photometric stability of the candidate standard DAWDs}\label{sec:valid}
In order to assess the 23 candidates as stable standards we monitored them 
by collecting time-series data with the Las Cumbres Observatory (LCO) 
network of telescopes (proposals LCO2016B-007, LCO2017AB-002, PI: Matheson). 

WDs can vary due to several reasons, depending on their 
effective temperature, atmosphere abundance and presence of 
magnetic activity or of an unseen faint companion star.

Hydrogen-rich atmosphere WDs might present gravity-mode
pulsations around $T_{eff}$ $\sim$ 12,000 K \citep[ZZ Ceti pulsators]{fontaine2008}. 
Our DAWDs were selected to have temperatures ($T_{eff} \gtrsim$ 20,000 K) 
outside the ZZ Ceti instability strip, so we do not expect them to be pulsators 
(note that SDSS-J172135.97+294016.0 will be removed from the network of standards since it 
has $T_{eff}$ = 9,261 K).
Strong magnetic fields can also cause flux variations in WDs
with a time scale from hours to days. 
These variations can be due to magnetically confined "spots" of
higher opacity modulating the stellar flux via stellar rotation \citep{dupuis2000, 
holberg2011}. Alternately, magnetic variations
can be due to to spots in the convective atmosphere \citep{brinkworth2004, brinkworth2013}.
However, the 23 candidate standard DAWDs have effective temperatures above
$\sim$ 20,000 K, and their atmosphere are fully radiative, so
they should not vary due to the presence of spots. 
Moreover, their spectra did not show Zeeman 
splitting of the Balmer lines indicative of the presence of a strong magnetic field (see \S 6).
On the other hand, the selected DAWDs could still vary due to the presence of an unseen
faint companion star, or to unknown factors, and we need to characterize
the amount of flux variation, if present, before setting these stars as spectrophotometric standards.
 A recent study by \citet{hermes2017}, based on precise \emph{Kepler} time-series
photometry, showed that $\sim$ 97\% of apparently isolated WDs are stable, or 
show less than 1\% flux variations, and they can still be used as spectrophotometric standards. 
Hermes et al.\ sample included mostly DA WDs but also several helium- or carbon-dominated
atmosphere WDs, with temperatures hotter than $\sim$ 8,000 K.

On the basis of the criteria used to select the 23 DAWDs and previous studies,
we do not expect a large fraction of our candidate spectrophotometric standards to vary. 
However, these 23 DAWDs have not yet been subject to a consistent and well-defined 
observational campaign to demonstrate a lack of variability at a wide range of time scales. 
WFC3 observations are obtained within a short time frame for each target, and so they are unsuitable as tests of variation. 
Prior ground-based surveys (SDSS, PanSTARRS) also do not have the necessary temporal coverage, 
and \emph{GAIA} does not provide variability constraints on these stars yet.

\subsection{Time-series photometry}\label{sec:series}
LCO observations consist of a sequence of geometrically spaced exposures in the Sloan $g$ filter, ranging from minutes to 
month-long time scales. A minimum of 20 exposures for each target were collected, spread over 2-3 months at different time intervals, for a total of about 800 images. 

Point-Spread Function (PSF) photometry was performed with DAOPHOTIV/ALLSTAR \citep{stetson1987} and ALLFRAME \citep{stetson1994}. 
The average FWHM for each frame was measured by using Source Extractor to exclude observations 
affected by poor observing conditions or bad focus, and these handful of images were excluded from the analysis.
All the exposures for each target were flux scaled to the best image, defined by the frame 
with the smallest average FWHM, which was used as a reference image. 
Light curves were then produced for each of the 23 targets.

In order to select candidate variables we used the \citet{welchstet1993} variability index $W$:

\begin{equation}
W =\sqrt{\frac{1}{n(n-1)}} \sum_{i=1}^{n}\frac{m_{i } - \bar{m}}{\sigma _{i}}
\end{equation}

where $m_i$ are the individual measurements and
$\bar{m}$ is the mean weighted magnitude of each identified object,
and $n$ is the total number of frames. The Welch-Stetson variability index was calculated for all the stars (from $\sim$ 500 to 1,000)
in the field of view. A sample of {\it stable} comparison stars was selected for each of the 23 DAWD observation.
This group of {\it stable} stars has a detection in every frame and a variability index, {\it var index}, $\le$ 1.2, 
the sharpness of the PSF in the range -0.5 $< sharpness <$ 0.5, to exclude extended objects
and CRs, and a proximity in magnitude to the target WD within $\sim$ 0.2 mag.

 An absolute calibration of the photometry is not performed since our goal is to demonstrate the lack of variability of the candidate standard DAWDs. 
However, we need to take into account spurious flux variations due to instrumental and atmospheric effects (observations are performed with 
different telescopes and detectors and from different sites in different conditions). The light curves of the selected {\it stable} stars are then compared to the light curves of the WDs in the same field. The variation around the mean of the {\it stable} star magnitudes was averaged 
and the average 1-$\sigma$ dispersion was estimated. This dispersion is used as a variability threshold for the systematic 
observational and instrumental effects.

\begin{figure*}
\begin{center}
\includegraphics[height=0.4\textheight,width=0.95\textwidth] {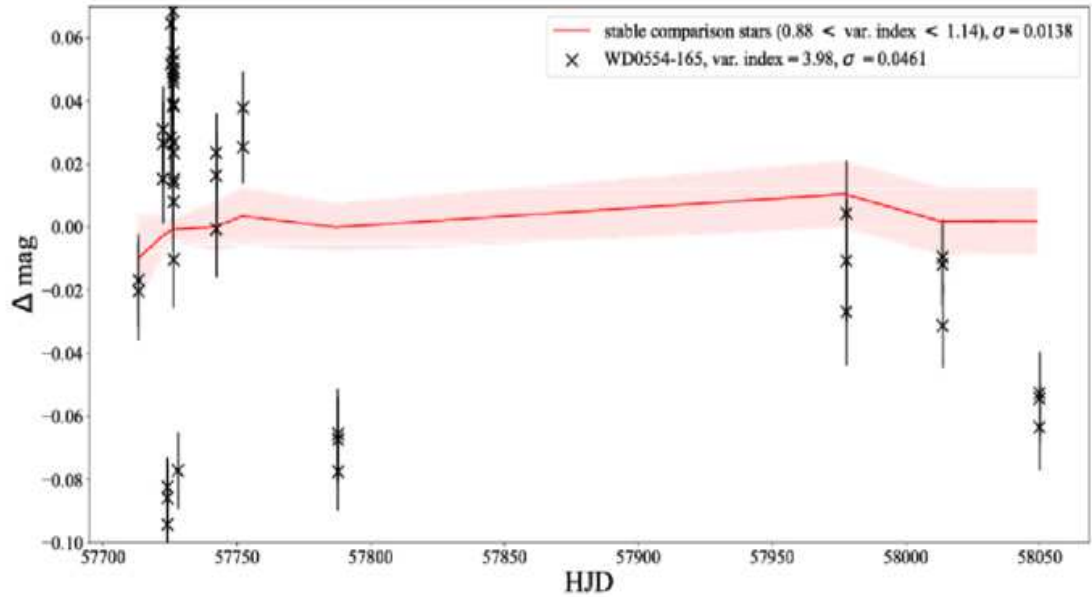} 
\caption{Single epoch minus the mean instrumental magnitude measurements for WD0554-165
as a function of observing epoch (black crosses). Averaged and binned relative magnitudes for a set of stable stars of comparable
instrumental magnitude in the same field of view are overplotted as a red shaded area. The variability index of the selected stars and the measurement
dispersions are listed. Error bars are shown. \label{fig:wd0554}}
\end{center}
\end{figure*}

\begin{figure*}
\begin{center}
\includegraphics[height=0.4\textheight,width=0.95\textwidth] {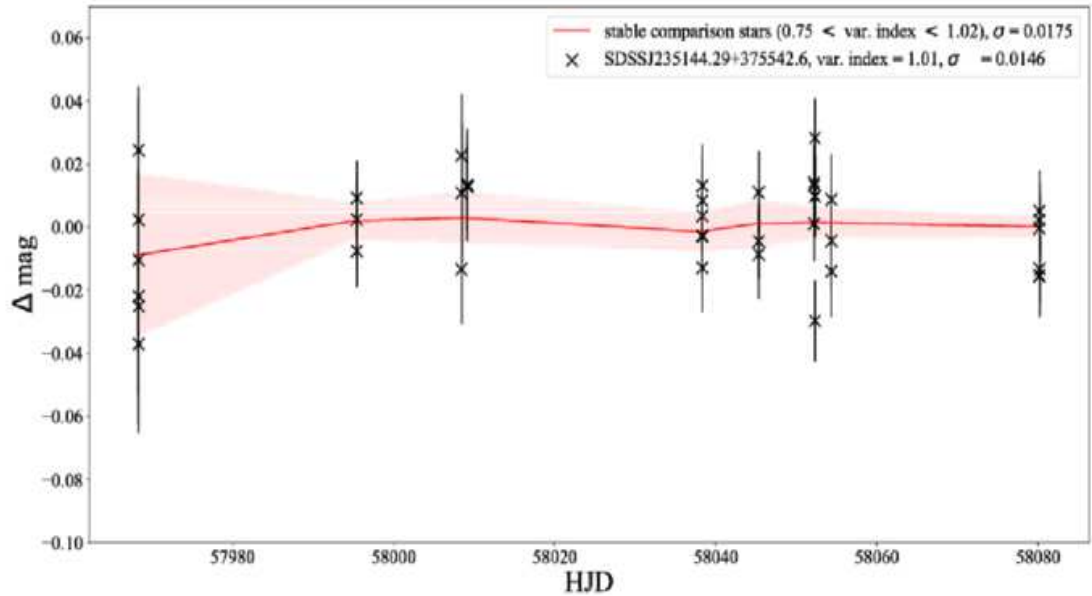} 
\caption{Same as Fig.~\ref{fig:wd0554} but for star SDSSJ235144.29+375542.6.\label{fig:sdssj235144}}
\end{center}
\end{figure*}

Fig.~\ref{fig:wd0554} shows the single epoch minus the weighted mean instrumental magnitude 
as a function of the Heliocentric Julian Date (HJD) for WD0554-165 (black crosses); averaged and binned relative magnitudes for 
a set of {\it stable} stars of comparable instrumental magnitude in the same field of view are also plotted as a red shaded area. 
The selected comparison stars have a variability index less than 1.2 while the WD0554-165 has a variability index of 3.98. 
WD0554-165 shows clear signs of variability, with variations of almost 0.2 mag and a measurement 1-$\sigma$ 
dispersion of $\sim$ 0.05 mag, compared to the {\it stable} star 
dispersion of  $\sigma \sim$ 0.01 mag.
Fig.~\ref{fig:sdssj235144} shows the same plot but for a stable DAWD, SDSSJ235144.29+375542.6:
its variability index is $\sim$ 1 and the dispersion of the measurement is $\sim$ 0.015 mag, smaller
than the measurement dispersions for the {\it stable} stars, $\sigma \sim$ 0.018 mag.

The light curve for SDSSJ20372.169-051302.964, a candidate binary system from spectroscopic data, shows variability 
with a dispersion of the measurements of $\sigma \sim$ 0.04 mag, a factor of 4 larger 
than the comparison star measurement dispersion, $\sigma \sim$ 0.01 mag, thus confirming its binary nature.

Other two stars in the sample, SDSSJ010322.10-002047.7 and SDSSJ102430.93-003207.0, show hints of variability but 
more and deeper exposures are needed to confirm these preliminary results.

Stars SDSSJ20372.169-051302.964 and WD0554-165 will be excluded from our network of spectrophotometric standard 
DAWDs due to their variable nature.

A more detailed analysis of the LCO photometry and the DAWD light curves will be presented in a forthcoming paper. We also plan to follow-up the candidate variable DAWDs with more observations from a larger ground-based telescope to understand the origin of their variability.

\section{Spectroscopic Observations}\label{sec:spectra}

Spectra of the DAWDs are used to determine $T_{eff}$ and
log~$g$.  These values are derived from the shape of the \ion{H}{1}
Balmer line profiles from H$\beta$ to H$\zeta$.  We flux-calibrated
the spectra to facilitate the analysis of the Balmer lines, but the
overall shape of the spectrum will retain uncertainties introduced by
the flux calibration process as well as the inherent uncertainty in
the standard stars used.  We emphasize that the spectral shapes are
not critical to the ultimate analysis of these DAWDs as
spectrophotometric standards.  The deviation of the calibrated spectral shape from a model spectrum is treated as a nuisance parameter in the fitting process, so minimizing the calibration error does improve the uncertainties in the end result.  Nonetheless, it is the values of $T_{eff}$ and log~$g$ from the Balmer lines in concert with the photometry that provides the ultimate calibration of these stars.

We used two different facilities to obtain spectra of our standard
star candidates.  As part of the \emph{HST} photometry proposal, we
were awarded Gemini time.  This amounted to 43 hours from Cycle 20
(split over Gemini semesters 2013A and 2013B) and 28.1 hours from
Cycle 22 (split over Gemini semesters 2015A and 2015B).  For most of
the time, we used Gemini South, but we also used Gemini North for the
northern targets.  At each site, we used the Gemini Multi-Object
Spectrograph \citep[GMOS,][]{hook04} in queue mode with a long slit to
obtain the spectra. For the 2013A and 2013B semesters, we used the
1\farcs5 slit, while for 2015A and 2015B, we used the 1\farcs0 slit.
The three GMOS detectors are not contiguous, so we used two different
grating tilts to fill in the inter-chip gaps.  The final spectra are
continuous from 3500~\AA\ to 6360~\AA with a dispersion of
0.92~\AA/pixel.  The resolution of the spectra is a function of the
seeing at the time of observation given the relatively wide slit
and the generally good seeing conditions at the Gemini sites.
Determining the resolution is an element of the data analysis process
that will be described in a later paper.

We found that the Gemini data were generally not of sufficiently high
quality for our purpose. The throughput of the GMOS system in the
blue is poor. In addition, standard stars and other calibrations were
frequently not obtained in conjunction with the spectra of the white
dwarfs. Finally, despite our request, the observations were typically not obtained at the
parallactic angle \citep{filippenko82} so slit losses resulting from
atmospheric dispersion resulted in compromised shapes for the spectral
energy distributions of the stars.  Because of these issues, we
instituted a program at the MMT Observatory to obtain alternate
spectra of our DAWDs.

At the MMT, we used the Blue Channel spectrograph \citep{schmidt89}
with the 300 line/mm grating.  We had a total of four successful
observing nights spread over three epochs.  For most of the
observations, we used the 1\farcs0 slit, but with the 1\farcs25 slit
for one epoch.  The wavelength coverage runs from 3400\AA\ to 8400\AA\
with a dispersion of 1.95~\AA/pixel.  All observations were obtained
at the parallactic angle and standard stars were observed on the same night.  As with the GMOS data, the resolution of the spectra depends
on the seeing at the time of observation.

\begin{figure*}
\hspace{-1.0cm}
\includegraphics[height=0.6\textheight,width=1.1\textwidth]{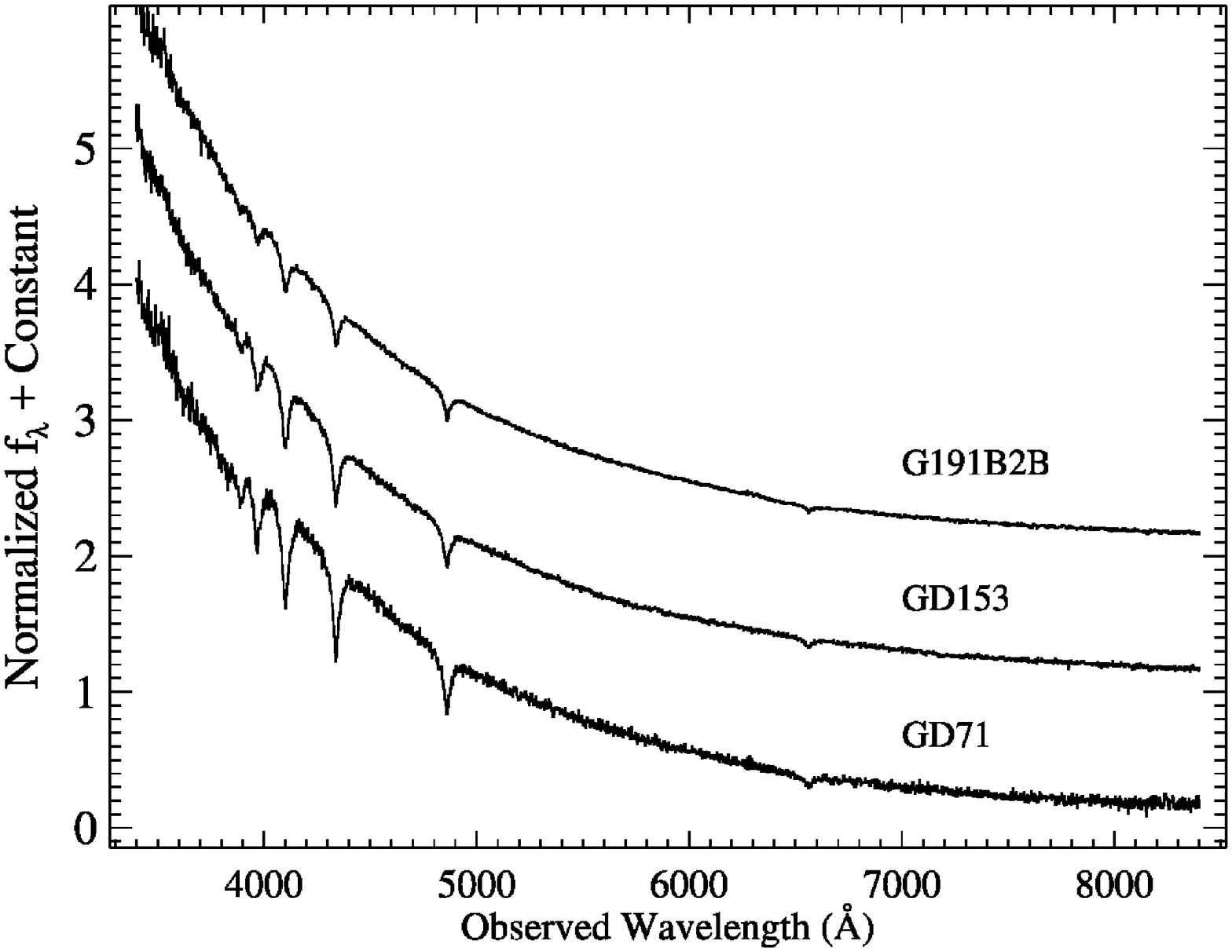} 
\caption{Spectra of the three \emph{HST} primary CALSPEC standards. \label{fig:HST} }
\end{figure*}

\begin{figure*}
\begin{center}
\hspace{-2.0cm}
\includegraphics[height=0.95\textheight,width=1.1\textwidth]{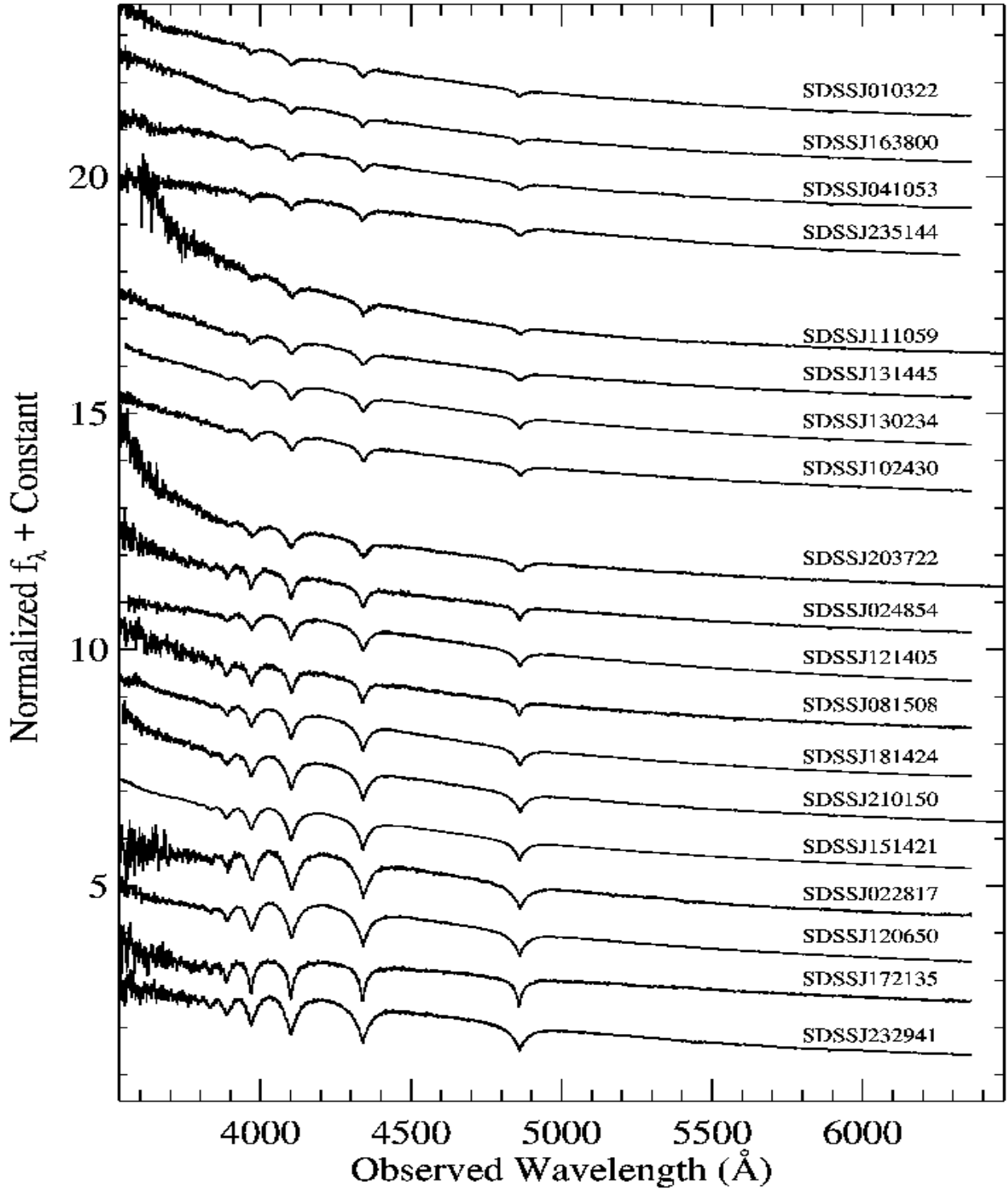} 
\caption{Spectra of the DAWD stars in our program obtained with
the GMOS instruments at Gemini.  Note that the unusual shapes of
some of the spectra are caused by atmospheric dispersion
effects and slit losses. Spectra are ordered by $T_{eff}$, with the hottest stars at the top.\label{fig:montgem} }
\end{center}
\end{figure*}

\begin{figure*}
\centering
\hspace{-2.0cm}
\includegraphics[height=0.95\textheight,width=1.1\textwidth]{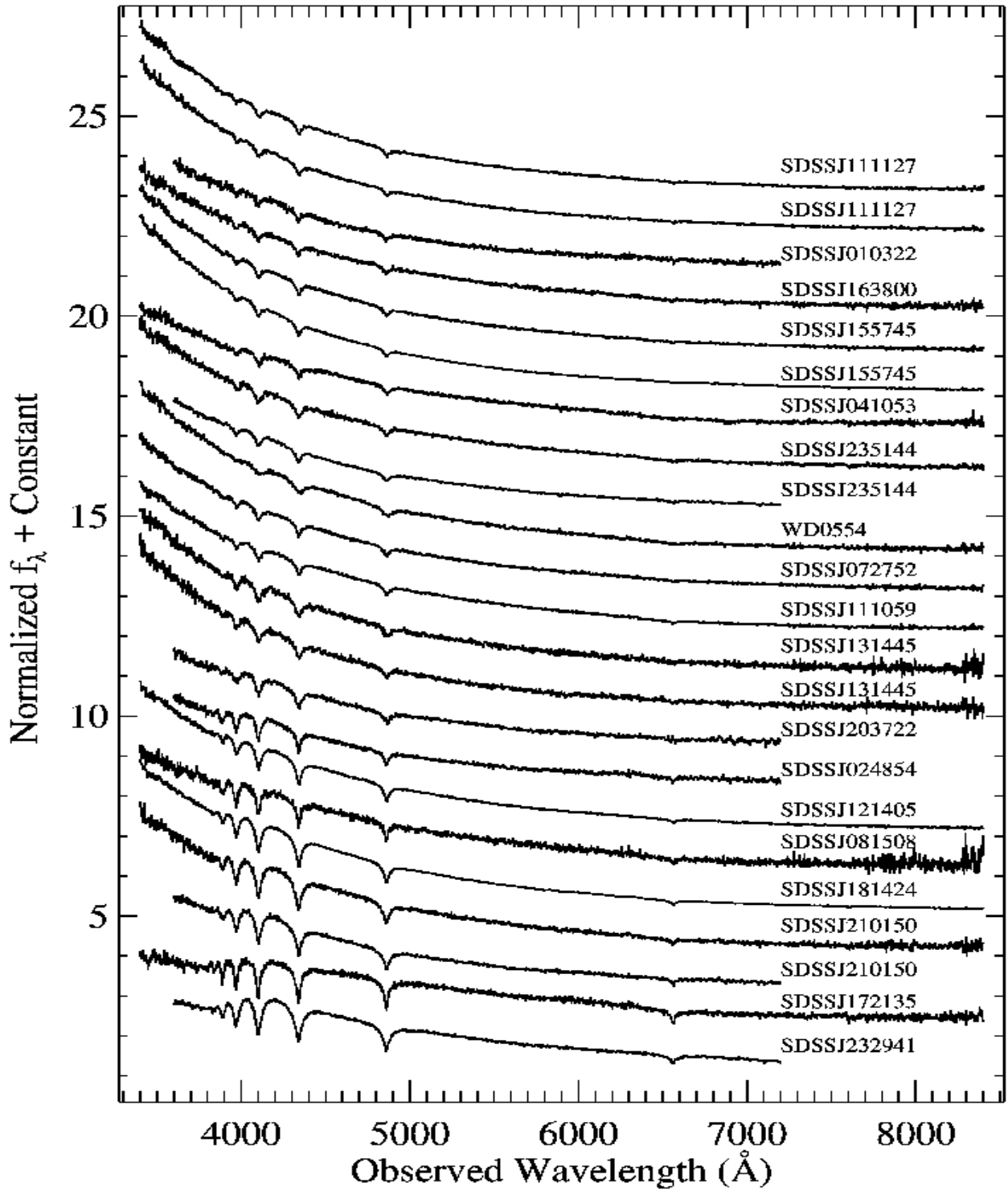}
\caption{Spectra of the WD stars in our program obtained with
the Blue Channel spectrograph at the MMT.  Spectra are ordered by
$T_{eff}$, with the hottest stars at the top.}
\label{fig:montmmt} 
\end{figure*}

Details of the observations with both facilities are presented in
Table~\ref{table:5}.  We used standard IRAF\footnote{IRAF is distributed by the National Optical Astronomy Observatory, which is operated by AURA under cooperative agreement with the NSF.} routines to process the CCD data and optimally extract \citep{horne86} the spectra.  The wavelength scale was evaluated via polynomial fits to calibration lamp spectra and then we resampled the WD spectra onto a linear scale with 1\AA/pixel and 2\AA/pixel for the GMOS and MMT data, respectively.  We used our own custom IDL routines to flux calibrate the data \citep{matheson08}.  Standard stars for each spectrum are listed in Table \ref{table:5}.  The spectra of our DAWDs are shown in three figures.  Fig.~\ref{fig:HST} shows the spectra of the three primary CALSPEC standards.  Spectra of stars obtained at Gemini are shown in Fig.~\ref{fig:montgem} while those obtained at the MMT are shown in Fig.~\ref{fig:montmmt}.  The details of the determination of $T_{eff}$ and log~$g$ will be described in a forthcoming analysis paper (Narayan et al.\ 2018, in prep.).

One of the WD stars showed indications of abnormality in
its spectra.  SDSSJ20372.169-051302.964 was observed over several nights with GMOS-S. There is a narrow emission feature present in the cores of the Balmer absorption lines. The emission feature moves relative to the broader line.  This may be the result of a low-luminosity companion or some other activity associated with the WD. This star is thus unsuitable for use as a spectrophotometric standard as the model spectra only apply to single, inactive DAWDs.

\begin{deluxetable*}{lccccccccc}
\tabletypesize{\scriptsize}
\tablecaption{List of properties of the \emph{HST} primary CALSPEC standards and the 23 candidate spectrophotometric standard DAWDs.\label{table:1} }   
\tablehead{
\colhead{Star}&
\colhead{Alt name}&
\colhead{RA\tablenotemark{a}}&
\colhead{DEC\tablenotemark{a}}&
\colhead{$PM_{RA}$\tablenotemark{a}}&
\colhead{$PM_{DEC}$\tablenotemark{a}}&
\colhead{SType\tablenotemark{a}}&
\colhead{Distance\tablenotemark{a}}&
\colhead{$T_{eff}$\tablenotemark{b}}&
\colhead{$log(g)$\tablenotemark{b}}\\ 
\colhead{}&   
\colhead{}& 
\colhead{(hh:mm:ss.s)}&
\colhead{(dd:mm:ss.s)}&
\colhead{(mas/yr)}&
\colhead{(mas/yr)}&
\colhead{}&
\colhead{(pc)}&
\colhead{(K)}&
\colhead{} }
\startdata 
G191B2B                       &  BD+52 913               & 5:05:30.613  & 52:49:51.956 &  12.592$\pm$0.122 &  -93.525$\pm$0.106 & DA.89  &  52.9$\pm$0.2  & 57340 & 7.48 \\
GD71                          &  GD71                    & 5:52:27.614  & 15:53:13.751 &  76.841$\pm$0.131 & -172.944$\pm$0.104 & DA1.5  &  51.9$\pm$0.2  & 32780 & 7.83 \\
GD153                         &  GD153                   & 12:57:02.337 & 22:01:52.68  & -38.410$\pm$0.119 & -202.953$\pm$0.116 & DA1.3  &  68.6$\pm$0.3  & 39390 & 7.77 \\
SDSSJ010322.19-002047.7       &  SDSSJ010322.19-002047.7 & 1:03:22.191  & -0:20:47.731  &   6.216$\pm$0.957 &   -6.313$\pm$0.549 & DA.67  &  1097$\pm$611  & 75000 & 7.81 \\
SDSSJ022817.16-082716.4       &  WD0225-086              & 2:28:17.169  & -8:27:16.409 &  12.241$\pm$1.461 &   3.827$\pm$1.000 & DA2.45 &  525$\pm$181   & 20555 & 7.87 \\
SDSSJ024854.96+334548.3       &  SDSSJ024854.96+334548.3 & 2:48:54.967  & 33:45:48.33  &   3.635$\pm$0.700 &   -4.718$\pm$0.442 & DA1.46 &  630$\pm$128   & 34497 & 7.30 \\
SDSSJ041053.632-063027.580    &  WD0408-066              & 4:10:53.634  & -6:30:27.749 &   8.620$\pm$0.411 &   9.700$\pm$0.237 & DA.77  &  1833$\pm$1248 & 65796 & 7.52 \\
WD0554-165                    &   WD0554-165             & 5:57:01.296  & -16:35:12.12 &  -7.188$\pm$0.399 &   4.781$\pm$0.623 & ƒ.	&  239$\pm$13	 &\ldots & \ldots \\
SDSSJ072752.76+321416.1       &  SDSSJ072752.76+321416.1 & 7:27:52.76   & 32:14:16.141 & -13.095$\pm$0.366 &   -7.094$\pm$0.373 & DA.88  &  990$\pm$198   & 57865 & 7.61 \\
SDSSJ081508.78+073145.7       &  SDSSJ081508.78+073145.7 & 8:15:08.779  & 7:31:45.804  &   3.135$\pm$1.384 &   0.313$\pm$0.794 & DA1.55 & \ldots	 & 32387 & 6.81 \\
SDSSJ102430.93-003207.0       &  SDSSJ102430.93-003207.0 & 10:24:30.932 & 0:32:07.03   &   -24.0	   &    -5.0	       & DA1.21 & \ldots	 & 41584 & 7.77 \\
SDSSJ111059.42-170954.2       &  SDSSJ111059.42-170954.2 & 11:10:59.428 & -17:09:54.27 &   5.045$\pm$0.418 &   -7.763$\pm$0.293 & DA.96  &  1333$\pm$359  & 52555 & 7.73 \\
SDSSJ111127.30+395628.0       &  SDSSJ111127.30+395628.0 & 11:11:27.309 & 39:56:28.079 &   3.277$\pm$0.445 &   3.095$\pm$0.598 & DA.75  &  648$\pm$166   & 67380 & 7.80 \\
SDSSJ120650.504+020143.810    &  WD1204+023              & 12:06:50.408 & 2:01:42.46   &  -4.594$\pm$0.663 &  -23.143$\pm$0.319 & DA2.02 &  590$\pm$130   & 24926 & 7.98 \\
SDSSJ121405.11+453818.5       &  CSO1291                 & 12:14:05.112 & 45:38:18.56  &   0.291$\pm$0.140 &   13.803$\pm$0.170 & DA1.43 &  495$\pm$31    & 35245 & 7.91 \\
SDSSJ130234.43+101238.9       &  SDSSJ130234.43+101238.9 & 13:02:34.441 & 10:12:39.01  & -12.523$\pm$0.252 &  -17.372$\pm$0.191 & DA1.20 &  389$\pm$17    & 42070 & 7.91 \\
SDSSJ131445.050-031415.588    &   WD1312-029             & 13:14:45.05  & -3:14:15.641 &  -4.102$\pm$1.190 &   -6.354$\pm$0.606 & DA1.05 &  1154$\pm$834  & 47818 & 7.76 \\
SDSSJ151421.27+004752.8       &  LB 769                  & 15:14:21.28  & 0:47:52.883  &   4.400$\pm$0.175 &  -27.041$\pm$0.222 & DA1.74 &  157$\pm$3	 & 28999 & 7.81 \\
SDSSJ155745.40+554609.7       &  WD1556+559              & 15:57:45.404 & 55:46:09.75  & -11.545$\pm$0.260 &  -21.340$\pm$0.204 & DA.79  &  688$\pm$52.   & 64122 & 7.65 \\
SDSSJ163800.360+004717.822    &  WD1635+008              & 16:38:00.366 & 0:47:17.801  &  -9.582$\pm$0.782 &   -2.797$\pm$0.467 & DA.77  &  876$\pm$283   & 65116 & 7.37 \\
SDSSJ172135.97+294016.0       &  SDSSJ172135.97+294016.0 & 17:21:35.981 & 29:40:15.996 & -21.454$\pm$0.564 &   10.452$\pm$0.638 & DA5.44 &  271$\pm$26    & 9261  & 8.33 \\
SDSSJ181424.075+785403.048    &  WD1817+788              & 18:14:24.122 & 78:54:02.909 & -11.041$\pm$0.103 &   11.292$\pm$0.132 & DA1.6  &  257$\pm$3	 & 31500 & 7.81 \\
SDSSJ20372.169-051302.964    &  WD 2034-053             & 20:37:22.167 & -5:13:03.029 &   3.106$\pm$0.647 &   -2.723$\pm$0.389 & DA1.33 &  912$\pm$324   & 37923 & 7.92 \\
SDSSJ210150.65-054550.9       & WD2059-059               & 21:01:50.657 & -5:45:50.969 &  10.828$\pm$0.456 &  -11.727$\pm$0.372 & DA1.75 &  662$\pm$107   & 28816 & 7.78 \\
SDSSJ232941.330+001107.755    &  WD2327-000              & 23:29:41.325 & 0:11:07.8    &  -8.299$\pm$0.384 &  -14.421$\pm$0.277 & DA2.37 &  318$\pm$25    & 21266 & 7.88 \\
SDSSJ235144.29+375542.6       & SDSSJ235144.29+375542.6  & 23:51:44.293 & 37:55:42.661 & -16.575$\pm$0.294 &  -10.048$\pm$0.185 & DA.95  &  765$\pm$134   & 53333 & 7.72 \\
\enddata
\tablenotetext{a}{Coordinates, proper motions, spectral type and distance measurements are from GAIA DR2.}
\tablenotetext{b}{Effective temperature and surface gravity measurements are from the SDSS or the Villanova catalogs.}
\end{deluxetable*}


\begin{deluxetable*}{lcccccccccc}
\tabletypesize{\scriptsize}
\tablecaption{Pan-STARRS1 photometry for the candidate spectrophotometric standard DA white dwarfs.\label{table:2}} 
\tablehead{
\colhead{Star}&
\colhead{RA}&
\colhead{DEC}&
\colhead{g}&
\colhead{$err_g$}&
\colhead{r}&
\colhead{$err_r$}&
\colhead{i}&
\colhead{$err_i$}&
\colhead{z}&
\colhead{$err_z$}\\
\colhead{}&
\colhead{(hh:mm:ss.s)}&
\colhead{(dd:mm:ss.s)}&
\colhead{(mag)}&
\colhead{(mag)}&
\colhead{(mag)}&
\colhead{(mag)}&
\colhead{(mag)}&
\colhead{(mag)}&
\colhead{(mag)}&
\colhead{(mag)}
}
\startdata               
G191B2B		             &  5:05:30.613  & 52:49:51.956  & $\ldots$	 & $\ldots$ & $\ldots$ & $\ldots$ & $\ldots$ & $\ldots$ & $\ldots$ & $\ldots$ \\
GD71			     &  5:52:27.614  & 15:53:13.751  &   12.846  &  0.003   &  13.284  &  0.001   &  13.629  & 0.005  & 13.921  & 0.003	\\
GD153  		             &  12:57:02.337 & 22:01:52.68   &   13.115  &  0.004   &  13.586  &  0.001   &  13.968  & 0.006  & 14.257  & 0.002	\\
SDSSJ010322.19-002047.7      &  1:03:22.191  & -0:20:47.731   &   19.093  &  0.010   &  19.570  &  0.019   &  19.979  & 0.017  & 20.130  & 0.064	\\
SDSSJ022817.16-082716.4      &  2:28:17.169  & -8:27:16.409  &   19.837  &  0.014   &  20.188  &  0.053   &  20.523  & 0.036  & 20.803  & 0.117	\\
SDSSJ024854.96+334548.3      &  2:48:54.967  & 33:45:48.33   &   18.351  &  0.007   &  18.699  &  0.006   &  18.972  & 0.012  &	19.198  & 0.031 \\
SDSSJ041053.632-063027.580   &  4:10:53.634  & -6:30:27.749  &   18.871  &  0.008   &  19.224  &  0.010   &  19.429  & 0.022  & 19.342  & 0.028	\\
WD0554-165                   &  5:57:01.296  & -16:35:12.12  &   17.787  &  0.012   &  18.237  &  0.012   &  18.628  & 0.010  & 18.916  & 0.010  \\
SDSSJ072752.76+321416.1      &  7:27:52.76   & 32:14:16.141  &   18.018  &  0.010   &  18.475  &  0.011   &  18.806  & 0.012  & 19.127  & 0.024 \\
SDSSJ081508.78+073145.7      &  8:15:08.779  & 7:31:45.804   &   19.781  &  0.040   &  20.328  &  0.037   &  20.625  & 0.073  & 20.710  & 0.165  \\
SDSSJ102430.93-003207.0      &  10:24:30.932 & 0:32:07.03    &   18.885  &  0.009   &  19.292  &  0.023   &  19.440  & 0.098  & 19.758  & 0.031  \\
SDSSJ111059.42-170954.2      &  11:10:59.428 & -17:09:54.27  &   17.895  &  0.005   &  18.302  &  0.009   &  18.607  & 0.015  & 18.957  & 0.026  \\
SDSSJ111127.30+395628.0      &  11:11:27.309 & 39:56:28.079  &   18.412  & 0.015    &  18.886  &  0.011   &  19.260  & 0.011  & 19.586  & 0.016  \\
SDSSJ120650.504+020143.810   &  12:06:50.408 & 2:01:42.46    &   18.693  & 0.010    &  19.096  &  0.029   &  19.388  & 0.024  & 19.645  & 0.034  \\
SDSSJ121405.11+453818.5      &  12:14:05.112 & 45:38:18.56   &   17.779  & 0.005    &  18.236  &  0.007   &  18.570  & 0.010  & 18.849  & 0.017  \\
SDSSJ130234.43+101238.9      &  13:02:34.441 & 10:12:39.01   &   17.052  & 0.003    &  17.494  &  0.003   &  17.858  & 0.006  & 18.114  & 0.009  \\
SDSSJ131445.050-031415.588   &  13:14:45.05  & -3:14:15.641  &   19.078  & 0.014    &  19.556  &  0.021   &  19.887  & 0.040  & 20.240  & 0.069  \\
SDSSJ151421.27+004752.8      &  15:14:21.28  & 0:47:52.883   &   15.720  & 0.002    &  16.101  &  0.004   &  16.434  & 0.002  & 16.715  & 0.005  \\
SDSSJ155745.40+554609.7      &  15:57:45.404 & 55:46:09.75   &   17.487  & 0.005    &  17.958  &  0.007   &  18.356  & 0.005  & 18.647  & 0.011  \\
SDSSJ163800.360+004717.822   &  16:38:00.366 & 0:47:17.801   &   18.860  & 0.013    &  19.314  &  0.022   &  19.611  & 0.013  & 19.816  & 0.053  \\
SDSSJ172135.97+294016.0      &  17:21:35.981 & 29:40:15.996  &   19.637  & 0.015    &  19.636  &  0.015   &  19.754  & 0.024  & 19.995  & 0.101  \\
SDSSJ181424.075+785403.048   &  18:14:24.122 & 78:54:02.909  &   16.573  & 0.005    &  17.007  &  0.003   &  17.358  & 0.004  & 17.651  & 0.009  \\
SDSSJ20372.169-051302.964   &  20:37:22.167 & -5:13:03.029  &   18.987  & 0.009    &  19.349  &  0.010   &  19.576  & 0.020  & 19.881  & 0.047  \\
SDSSJ210150.65-054550.9      &  21:01:50.657 & -5:45:50.969  &   18.652  & 0.009    &  19.052  &  0.008   &  19.410  & 0.018  & 19.703  & 0.033  \\
SDSSJ232941.330+001107.755   &  23:29:41.325 & 0:11:07.8     &   18.134  & 0.006    &  18.452  &  0.005   &  18.772  & 0.008  & 19.003  & 0.017  \\
SDSSJ235144.29+375542.6      &  23:51:44.293 & 37:55:42.661  &   18.085  & 0.004    &  18.447  &  0.013   &  18.776  & 0.010  & 19.100  & 0.036	 \\
\enddata
\end{deluxetable*}




\begin{deluxetable*}{lcccccccc}
\tabletypesize{\scriptsize}
\tablecaption{GAIA DR2 photometry for the candidate spectrophotometric standard DA white dwarfs. \label{table:3}}
\tablehead{
\colhead{Star}&
\colhead{RA}&
\colhead{DEC}&
\colhead{G}&
\colhead{$err_G$}&
\colhead{Rp}&
\colhead{$err_{Rp}$}&
\colhead{Bp}&
\colhead{$err_{Bp}$}\\
\colhead{}&
\colhead{(hh:mm:ss.s)}&
\colhead{(dd:mm:ss.s)}&
\colhead{mag}&
\colhead{mag}&
\colhead{mag}&
\colhead{mag}&
\colhead{mag}&
\colhead{mag}
}
\startdata                      
G191B2B		             & 5:05:30.613  & 52:49:51.956 & 11.738 & 0.001 & 12.067 & 0.002 & 11.487 & 0.015  \\
GD71			     & 5:52:27.614  & 15:53:13.751 & 13.026 & 0.002 & 13.299 & 0.002 & 12.77  & 0.012  \\
GD153  		             & 12:57:02.337 & 22:01:52.68  & 13.322 & 0.0   & 13.629 & 0.001 & 13.081 & 0.005  \\
SDSSJ010322.19-002047.7      & 1:03:22.191  & -0:20:47.731  & 19.356 & 0.004 & 19.577 & 0.072 & 19.154 & 0.03   \\
SDSSJ022817.16-082716.4      & 2:28:17.169  & -8:27:16.409 & 20.046 & 0.01  & 20.192 & 0.141 & 19.869 & 0.139  \\
SDSSJ024854.96+334548.3      & 2:48:54.967  & 33:45:48.33  & 18.561 & 0.003 & 18.704 & 0.031 & 18.333 & 0.047  \\
SDSSJ041053.632-063027.580   & 4:10:53.634  & -6:30:27.749 & 19.024 & 0.002 & 19.013 & 0.023 & 18.861 & 0.025  \\
WD0554-165                   & 5:57:01.296  & -16:35:12.12 & 17.98  & 0.003 & 18.306 & 0.03  & 17.726 & 0.022  \\
SDSSJ072752.76+321416.1      & 7:27:52.76   & 32:14:16.141 & 18.232 & 0.003 & 18.458 & 0.036 & 17.944 & 0.007  \\
SDSSJ081508.78+073145.7      & 8:15:08.779  & 7:31:45.804  & 19.996 & 0.005 & 20.278 & 0.166 & 19.695 & 0.044  \\
SDSSJ102430.93-003207.0      & 10:24:30.932 & 0:32:07.03   & 19.12  & 0.005 & 19.297 & 0.105 & 18.94  & 0.059  \\
SDSSJ111059.42-170954.2      & 11:10:59.428 & -17:09:54.27 & 18.089 & 0.002 & 18.347 & 0.02  & 17.852 & 0.011  \\
SDSSJ111127.30+395628.0      & 11:11:27.309 & 39:56:28.079 & 18.69  & 0.003 & 18.955 & 0.075 & 18.365 & 0.022  \\
SDSSJ120650.504+020143.810   & 12:06:50.408 & 2:01:42.46   & 18.885 & 0.002 & 18.957 & 0.03  & 18.651 & 0.017  \\
SDSSJ121405.11+453818.5      & 12:14:05.112 & 45:38:18.56  & 18.002 & 0.001 & 18.154 & 0.038 & 17.757 & 0.011  \\
SDSSJ130234.43+101238.9      & 13:02:34.441 & 10:12:39.01  & 17.268 & 0.001 & 17.527 & 0.012 & 17.044 & 0.006  \\
SDSSJ131445.050-031415.588   & 13:14:45.05  & -3:14:15.641 & 19.354 & 0.004 & 19.631 & 0.082 & 19.082 & 0.042  \\
SDSSJ151421.27+004752.8      & 15:14:21.28  & 0:47:52.883  & 15.905 & 0.001 & 16.119 & 0.005 & 15.743 & 0.009  \\
SDSSJ155745.40+554609.7      & 15:57:45.404 & 55:46:09.75  & 17.721 & 0.002 & 18.019 & 0.018 & 17.452 & 0.014  \\
SDSSJ163800.360+004717.822   & 16:38:00.366 & 0:47:17.801  & 19.065 & 0.002 & 19.313 & 0.04  & 18.853 & 0.019  \\
SDSSJ172135.97+294016.0      & 17:21:35.981 & 29:40:15.996 & 19.648 & 0.005 & 19.528 & 0.037 & 19.733 & 0.042  \\
SDSSJ181424.075+785403.048   & 18:14:24.122 & 78:54:02.909 & 16.773 & 0.002 & 17.031 & 0.007 & 16.57  & 0.009  \\
SDSSJ20372.169-051302.964   & 20:37:22.167 & -5:13:03.029 & 19.148 & 0.003 & 19.375 & 0.055 & 18.982 & 0.018  \\
SDSSJ210150.65-054550.9      & 21:01:50.657 & -5:45:50.969 & 18.867 & 0.002 & 19.095 & 0.044 & 18.654 & 0.021  \\
SDSSJ232941.330+001107.755   & 23:29:41.325 & 0:11:07.8    & 18.323 & 0.002 & 18.394 & 0.028 & 18.187 & 0.021  \\
SDSSJ235144.29+375542.6      & 23:51:44.293 & 37:55:42.661 & 18.272 & 0.002 & 18.417 & 0.014 & 18.056 & 0.016  \\
\enddata
\end{deluxetable*}



\begin{deluxetable*}{lccccccccc}
\tabletypesize{\scriptsize}
\tablecaption{Log of the observations collected with the Wide Field Camera 3 on board the Hubble Space Telescope
during cycles 20 and 22 (proposal IDs 12967 and 13711, PI: A. Saha). \label{table:4}}  
\tablehead{
\colhead{Star}&
\colhead{PID\tablenotemark{a}}&
\colhead{Image name}&
\colhead{RA}&
\colhead{DEC}&
\colhead{Filter}&
\colhead{Exposure time}&
\colhead{Date Obs.}&
\colhead{Time Obs.}&
\colhead{Aperture}\\
\colhead{}&  
\colhead{}&  
\colhead{}&  
\colhead{(hh:mm:ss.s)}&
\colhead{(dd:mm:ss.s)}&
\colhead{}& 
\colhead{(s)}&
\colhead{(YYYY/MM/DD)}& 
\colhead{(UT)}&
\colhead{}
}
\startdata
\multicolumn{10}{c}{Cycle 20} \\
\hline
SDSSJ010322.19-002047.7    & 12967 & ibyn01wxq & 01:03:22.1 & 00:20:47.7 & F336W & 160 & 2013-09-13 & 21:46:26 & UVIS1-FIX \\ 
SDSSJ010322.19-002047.7    & 12967 & ibyn01x0q & 01:03:22.1 & 00:20:47.7 & F336W & 160 & 2013-09-13 & 21:51:25 & UVIS1-FIX \\ 
SDSSJ010322.19-002047.7    & 12967 & ibyn01x2q & 01:03:22.1 & 00:20:47.7 & F336W & 160 & 2013-09-13 & 23:17:30 & UVIS1-FIX \\ 
SDSSJ010322.19-002047.7    & 12967 & ibyn01woq & 01:03:22.1 & 00:20:47.7 & F475W & 120 & 2013-09-14 & 20:13:32 & UVIS1-FIX \\ 
SDSSJ010322.19-002047.7    & 12967 & ibyn01xjq & 01:03:22.1 & 00:20:47.7 & F475W & 160 & 2013-09-14 & 01:08:35 & UVIS1-FIX \\ 
SDSSJ010322.19-002047.7    & 12967 & ibyn01xmq & 01:03:22.1 & 00:20:47.7 & F475W & 160 & 2013-09-14 & 01:13:33 & UVIS1-FIX \\ 
SDSSJ010322.19-002047.7    & 12967 & ibyn01wtq & 01:03:22.1 & 00:20:47.7 & F625W & 350 & 2013-09-13 & 20:30:47 & UVIS1-FIX \\ 
SDSSJ010322.19-002047.7    & 12967 & ibyn01x8q & 01:03:22.1 & 00:20:47.7 & F625W & 355 & 2013-09-13 & 23:36:41 & UVIS1-FIX \\ 
SDSSJ010322.19-002047.7    & 12967 & ibyn01wqq & 01:03:22.1 & 00:20:47.7 & F775W & 605 & 2013-09-13 & 20:18:08 & UVIS1-FIX \\ 
SDSSJ010322.19-002047.7    & 12967 & ibyn01x5q & 01:03:22.1 & 00:20:47.7 & F775W & 680 & 2013-09-13 & 23:22:47 & UVIS1-FIX \\ 
SDSSJ010322.19-002047.7    & 12967 & ibyn01wvq & 01:03:22.1 & 00:20:47.7 & F160W & 499 & 2013-09-14 & 21:36:26 & IR-FIX \\
SDSSJ010322.19-002047.7    & 12967 & ibyn01xhq & 01:03:22.1 & 00:20:47.7 & F160W & 499 & 2013-09-14 & 00:58:35 & IR-FIX \\ SDSSJ041053.632-063027.580 & 12967 & ibyn02lhq & 04:10:53.6 & -06:30:27.7 & F336W & 160 & 2013-08-27 & 22:02:46 & UVIS1-FIX \\ 
\enddata
\tablenotetext{a}{Program ID}
\tablecomments{Table \ref{table:4} is published in its entirety in the machine readable format.  A portion is
shown here for guidance regarding its form and content.}
\end{deluxetable*}


\begin{deluxetable*}{lcccrrccrr}
\tabletypesize{\scriptsize}
\tablecaption{Log of the spectroscopic observations. \label{table:5}}
\tablehead{\colhead{Star} & 
\colhead{UT Date} &
\colhead{Tel.\tablenotemark{a}} &
\colhead{Range}  &
\colhead{Res.\tablenotemark{b}} &
\colhead{P.A.\tablenotemark{c}} &
\colhead{Airmass} & 
\colhead{Flux Std.\tablenotemark{d}} &
\colhead{Slit} &
\colhead{Exposure} \\
\colhead{} &
\colhead{} &
\colhead{} &
\colhead{(\AA)} &
\colhead{(\AA)} &
\colhead{($^\circ$)} &
\colhead{} &
\colhead{} &
\colhead{($^{\prime\prime}$)} &
\colhead{(s)} }
\startdata
G191B2B                    &  2015-01-24 &         MMTO &  3400-8400 &  8 &  -111.8 &  1.2 &       Feige34 &      1.0 &      8 \\
GD153                      &  2015-05-18 &         MMTO &  3400-8400 &  8 &   -52.9 &  1.0 &     BD+284211 &      1.0 &     15 \\
GD71                       &  2015-01-24 &         MMTO &  3400-8400 &  8 &    12.8 &  1.0 &       Feige34 &      1.0 &     10 \\
SDSSJ010322.19-002047.7    &  2013-11-29 &     GEMINI-S &  3500-6360 & 10 &   180.0 &  1.2 &          GD71 &    1.5 &   6x1500 \\
SDSSJ010322.19-002047.7    &  2015-10-11 &         MMTO &  3400-8400 &  8 &    13.5 &  1.2 &     BD+284211 &     1.25 &   2x1200 \\
SDSSJ022817.16-082716.4    &  2013-10-23{\tablenotemark{X}} &     GEMINI-S &  3500-6360 & 10 &     0.0 &  1.2 &          GD71 &    1.5 &   7x1500 \\
SDSSJ024854.96+334548.3    &  2015-10-08{\tablenotemark{X}} &     GEMINI-N &  3520-6360 & 7 &   232.0 &  1.1 &     BD+284211/G191B2B &    1.0 &    8x999 \\
SDSSJ024854.96+334548.3    &  2015-10-11.5 &         MMTO &  3400-8400 &  8 &    92.4 &  1.1 &     BD+284211 &     1.25 &   5x1200 \\
SDSSJ041053.632-063027.580 &  2013-12-04{\tablenotemark{X}} &     GEMINI-S &  3500-6360 & 10 &   180.0 &  1.2 &          GD71 &    1.5 &   6x1500 \\
SDSSJ041053.632-063027.580 &  2015-01-24 &         MMTO &  3400-8400 &  8 &    12.8 &  1.3 &       Feige34 &      1.0 &    3x900 \\
WD0554-165                 &  2015-01-24 &         MMTO &  3400-8400 &  8 &    12.8 &  1.5 &       Feige34 &      1.0 &    3x900 \\
SDSSJ072752.76+321416.1    &  2015-01-24 &         MMTO &  3400-8400 &  8 &   -93.5 &  1.0 &       Feige34 &      1.0 &    3x900 \\
SDSSJ081508.78+073145.7    &  2013-07-07{\tablenotemark{X}} &     GEMINI-S &  3500-6360 & 12 &     0.0 &  1.3 &  GD71/Feige110 &    1.5 &   6x1500 \\
SDSSJ081508.78+073145.7    &  2015-01-24 &         MMTO &  3400-8400 &  8 &     9.3 &  1.1 &       Feige34 &      1.0 &    4x900 \\
SDSSJ102430.93-003207.0    &  2013-02-15 &     GEMINI-S &  3500-6360 & 12 &     0.0 &  1.3 &      Feige110 &    1.5 &   6x1500 \\
SDSSJ111059.43-170954.1    &  2015-01-24 &         MMTO &  3400-8400 &  8 &    -5.2 &  1.5 &       Feige34 &      1.0 &    3x900 \\
SDSSJ111059.43-170954.1    &  2015-05-18 &     GEMINI-S &  3500-6500 & 8 &     0.0 &  1.2 &       Feige67 &    1.0 &    8x700 \\
SDSSJ111127.30+395628.0    &  2015-01-24 &         MMTO &  3400-8400 &  8 &  -111.6 &  1.0 &       Feige34 &      1.0 &    3x900 \\
SDSSJ111127.30+395628.0    &  2015-05-18 &         MMTO &  3400-8400 &  8 &   130.0 &  1.0 &     BD+284211 &      1.0 &    2x900 \\
SDSSJ120650.504+020143.810 &  2013-03-10 &     GEMINI-S &  3500-6360 & 10 &    35.0 &  1.2 &      Feige110 &    1.5 &   6x1500 \\
SDSSJ121405.11+453818.5    &  2015-02-18 &     GEMINI-N &  3520-6360 & 8 &   130.0 &  1.2 &       Feige34 &    1.0 &    6x899 \\
SDSSJ121405.11+453818.5    &  2015-05-18 &         MMTO &  3400-8400 &  8 &  1000.0 &  1.0 &     BD+284211 &      1.0 &    3x900 \\
SDSSJ130234.43+101238.9    &  2013-02-15{\tablenotemark{X}} &     GEMINI-S &  3500-6360 & 10 &   138.0 &  1.4 &      Feige110 &    1.5 &   8x1200 \\
SDSSJ131445.050-031415.588 &  2013-03-09 &     GEMINI-S &  3500-6360 & 10 &   340.0 &  1.1 &      Feige110 &    1.5 &   6x1500 \\
SDSSJ131445.050-031415.588 &  2015-01-24 &         MMTO &  3400-8400 &  8 &   -22.4 &  1.3 &       Feige34 &      1.0 &    4x900 \\
SDSSJ131445.050-031415.588 &  2015-05-18 &         MMTO &  3400-8400 &  8 &     6.4 &  1.2 &     BD+284211 &      1.0 &    2x900 \\
SDSSJ151421.27+004752.8    &  2013-03-10{\tablenotemark{X}} &     GEMINI-S &  3500-6360 & 10 &     0.0 &  1.3 &      Feige110 &    1.5 &   8x1200 \\
SDSSJ155745.40+554609.7    &  2015-01-24 &         MMTO &  3400-8400 &  8 &  -113.9 &  1.2 &       Feige34 &      1.0 &    2x900 \\
SDSSJ155745.40+554609.7    &  2015-05-18 &         MMTO &  3400-8400 &  8 &  -129.8 &  1.1 &     BD+284211 &      1.0 &    4x900 \\
SDSSJ163800.360+004717.822 &  2013-04-08 &     GEMINI-S &  3500-6360 & 10 &     0.0 &  1.3 &      Feige110 &    1.5 &   6x1500 \\
SDSSJ163800.360+004717.822 &  2015-05-18 &         MMTO &  3400-8400 &  8 &    22.0 &  1.2 &     BD+284211 &      1.0 &    4x900 \\
SDSSJ172135.97+294016.0    &  2013-06-04 &     GEMINI-S &  3500-6360 & 10 &   180.0 &  2.1 &      Feige110 &    1.5 &   6x1500 \\
SDSSJ172135.97+294016.0    &  2015-05-18 &         MMTO &  3400-8400 &  8 &   -77.6 &  1.1 &     BD+284211 &      1.0 &    4x900 \\
SDSSJ181424.13+785402.9    &  2015-04-27 &     GEMINI-N &  3520-6360 & 8 &     0.0 &  2.0 &       Feige34 &    1.0 &    6x699 \\
SDSSJ181424.13+785402.9    &  2015-05-18 &         MMTO &  3400-8400 &  8 &  -152.3 &  1.5 &     BD+284211 &      1.0 &    3x900 \\
SDSSJ20372.169-051302.964  &  2014-07-14{\tablenotemark{X}} &     GEMINI-S &  3400-6500 & 10 &     0.0 &  1.1 &      Feige110 &    1.5 &   8x1500 \\
SDSSJ20372.169-051302.964  &  2015-10-12 &         MMTO &  3400-8400 &  8 &    -8.4 &  1.3 &     BD+284211 &     1.25 &   3x1200 \\
SDSSJ210150.65-054550.9    &  2014-07-20 &     GEMINI-S &  3400-6500 & 10 &     0.0 &  1.2 &      Feige110 &    1.5 &   6x1300 \\
SDSSJ210150.65-054550.9    &  2015-05-18 &         MMTO &  3400-8400 &  8 &   -39.9 &  1.5 &     BD+284211 &      1.0 &    2x900 \\
SDSSJ210150.65-054550.9    &  2015-10-11 &         MMTO &  3400-8400 &  8 &     1.7 &  1.3 &     BD+284211 &     1.25 &   6x1200 \\
SDSSJ232941.330+001107.755 &  2015-10-01{\tablenotemark{X}} &    GEMINI-N &  3520-6360 & 8 &    12.0 &  1.2 &     BD+284211 &    1.0 &   11x1099 \\
SDSSJ232941.330+001107.755 &  2015-10-11 &         MMTO &  3400-8400 &  8 &   -14.2 &  1.2 &     BD+284211 &     1.25 &   6x1200 \\
SDSSJ235144.29+355542.6    &  2015-05-18 &         MMTO &  3400-8400 &  8 &   -74.2 &  1.7 &     BD+284211 &      1.0 &    900+765 \\
SDSSJ235144.29+355542.6    &  2015-09-15 &     GEMINI-N &  3520-6360 & 8 &   180.0 &  1.2 &     BD+284211 &    1.0 &    6x999 \\
SDSSJ235144.29+355542.6    &  2015-10-11 &         MMTO &  3400-8400 &  8 &   126.4 &  1.0 &     BD+284211 &     1.25 &   5x1200 \\
\enddata
\tablenotetext{a}{Telescope used to obtain given spectrum.  GEMINI-N and GEMINI-S denote the use of
GMOS at either the northern or southern site for the Gemini Observatory.  MMTO denotes the use of
the Blue Channel spectrograph at the MMT Observatory.}
\tablenotetext{b}{Resolution of the spectrum as determined from the full-width at half maximum of sky
lines present in the two-dimensional spectrum.  This is, in general, an upper limit as the
resolution for the stellar spectrum is determined by the seeing and the slit width.  In many cases,
especially with the wider slits used with GMOS, the resolution of the spectrum is better than this
reported value.}
\tablenotetext{c}{The position angle of the observations.  Spectra from the MMT were typically
observed at the parallactic angle, while the Gemini data were not.}
\tablenotetext{d}{Flux standards used to calibrate the data: Feige~34, BD+28$^{\circ}$4211, Feige~110,
---\citep{stone77, massey88, massey90}; GD71---\citep{bohlin95}; G191B2B---\citep{oke74, massey88}.}
\tablenotetext{e}{Spectra from Gemini were obtained in queue mode and thus could be observed over
multiple nights.  The UT date reported for these stars is represents an average of the actual
dates. Note that observations for SDSSJ081508 were separated by 10 months.}
\end{deluxetable*}


\begin{deluxetable}{lccccc}
\tabletypesize{\scriptsize}
\tablecaption{Synthetic magnitudes and fluxes in the AB and ST photometric system for the 
three \emph{HST} primary CALSPEC DAWDs as simulated by using {\it Pysynphot}. The pivot wavelength for 
each filter is also listed. See text for more details.\label{table:6}}
\tablehead{
\colhead{Filter}&
\colhead{$\lambda_p$}&
\colhead{AB mag}&
\colhead{$F_{\nu}$}&
\colhead{ST mag}& 
\colhead{$F_{\lambda}$}\\
\colhead{}&
\colhead{\AA}&
\colhead{(mag)}&
\colhead{(erg cm$^{-2}$ s$^{-1}$ Hz$^{-1}$)}&
\colhead{(mag)}&
\colhead{(erg cm$^{-2}$ s$^{-1}$ Hz$^{-1}$)}
}
\startdata
\multicolumn{6}{c}{GD153} \\
\hline
F275W &  2,703  & 12.200 &  4.78e-25 & 10.669 & 1.96e-13  \\
F336W &  3,354  & 12.566 &  3.41e-25 & 11.503 & 9.09e-14   \\
F475W & 4,770   & 13.098 &  2.09e-25 & 12.799 & 2.76e-14   \\
F625W & 6,240   & 13.598 &  1.32e-25 & 13.882 & 1.02e-14  \\
F775W & 7,651   & 14.004 &  9.09e-26 & 14.730 & 4.66e-15   \\
F160W &15,369  & 15.414 &  2.48e-26 & 17.654 & 3.15e-16  \\
\hline
\multicolumn{6}{c}{GD71} \\
\hline
F275W & 2,703  & 11.981 & 5.85e-25 & 10.450 & 2.40e-13  \\
F336W & 3,354  & 12.327 & 4.26e-25 & 11.264  & 1.13e-13   \\
F475W & 4,770  & 12.794 & 2.77e-25 & 12.496  & 3.64e-14  \\
F625W & 6,240  & 13.275 & 1.78e-25 & 13.558  & 1.37e-14  \\
F775W & 7,651  & 13.672 & 1.23e-25  & 14.398 & 6.32e-15   \\
F160W & 15,369 & 15.060 & 3.43e-26  & 17.301 & 4.36e-16  \\
\hline
\multicolumn{6}{c}{G191B2B} \\
\hline 
F275W & 2,703  & 10.492 &  2.30e-24 & 8.960  & 9.46e-13   \\
F336W & 3,354  & 10.892 &  1.60e-24 & 9.829  & 4.25e-13  \\
F475W & 4,770  & 11.500 &  9.12e-25 & 11.201 & 1.20e-13  \\
F625W & 6,240  & 12.030 &  5.60e-25 & 12.314 & 4.31e-14  \\
F775W & 7,651  & 12.449 &  3.81e-25 & 13.175 & 1.95e-14  \\
F160W & 15,369 & 13.885 &  1.01e-25 & 16.125 & 1.29e-15  \\
\enddata
\end{deluxetable}



\begin{deluxetable*}{lcccccc}
\tabletypesize{\scriptsize}
\tablecaption{Zero points and their uncertainties for WFC3-UVIS (aperture radius $r =$ 7.5 pixels) and IR ($r =$ 5 pixels) 
observations in the AB photometric system. Zero points are derived by using observations of the 
three \emph{HST} primary CALSPEC standards as measured with three different methods. 
\label{table:7}}
\tablehead{
\colhead{Filter}&
\colhead{ZP (DAOPHOT)}&
\colhead{eZP (DAOPHOT)}&
\colhead{ZP (ILAPH)}&
\colhead{eZP (ILAPH)}&
\colhead{ZP (Sextractor)}&
\colhead{eZP (Sextractor)}\\
\colhead{}&
\colhead{(mag)}&
\colhead{(mag)}&
\colhead{(mag)}&
\colhead{(mag)}&
\colhead{(mag)}&
\colhead{(mag)}
}
\startdata
$F275W$   &   24.0612 &  0.0008 & 24.0596 & 0.0009 & 24.0594 & 0.0009 \\
$F336W$   &   24.5910 &  0.0008 & 24.5899 & 0.0008 & 24.5889 & 0.0010 \\
$F475W$   &   25.5780 &  0.0007 & 25.5774 & 0.0009 & 25.5761 & 0.0007 \\
$F625W$   &   25.4073 &  0.0005 & 25.4056 & 0.0007 & 25.4043 & 0.0005 \\
$F775W$   &   24.7207 &  0.0006 & 24.7189 & 0.0008 & 24.7171 & 0.0006 \\
$F160W$   &   25.8092 &  0.0007 & 25.8116 & 0.0009 & 25.8106 & 0.0007 \\
\enddata
\end{deluxetable*}



\begin{deluxetable}{lccccccc}
\tabletypesize{\scriptsize}
\tablecaption{Zero points and their uncertainties for 10 pixel aperture radius (WFC3-UVIS) and infinity (WFC3-UVIS, WFC3-IR)
in the AB photometric system.
Zero points are derived by using our observations of the three \emph{HST} primary CALSPEC standards 
as measured with DAOPHOT (first columns) and the official WFC3 values are in the last two columns\tablenotemark{a}. \label{table:8}}
\tablehead{
\colhead{Filter}&
\colhead{ZP$_{10}$} &
\colhead{eZP$_{10}$} &
\colhead{ZP$_{inf}$} &
\colhead{eZP$_{inf}$} &
\colhead{ZP$_{10}$(WFC3)} &
\colhead{ZP$_{inf}$(WFC3)} \\
\colhead{}&
\colhead{(mag)}&
\colhead{(mag)}&
\colhead{(mag)}&
\colhead{(mag)}&
\colhead{(mag)}&
\colhead{(mag)}
}
\startdata
$F275W$   &  24.0853 &  0.0007 & 24.2293 & 0.0008   & 24.075  & 24.224   \\
$F336W$   &  24.6131 &  0.0008 & 24.7344 & 0.0008   & 24.608  & 24.734 \\
$F475W$   &  25.6034 &  0.0006 & 25.7057 & 0.0007   & 25.604  & 25.709  \\
$F625W$   &  25.4310 &  0.0005 & 25.5334 & 0.0005   & 25.427  & 25.532 \\
$F775W$   &  24.7522 &  0.0005 & 24.8571 & 0.0006   & 24.753  & 24.859  \\
$F160W$   &  $\ldots$  & $\ldots$ & 25.9580 & 0.0007   & $\ldots$ & 25.946  \\
\enddata
\tablenotetext{a}{Current WFC3 UVIS and IR official ZPs can be found at
http://www.stsci.edu/hst/wfc3/phot\_zp\_lbn}
\end{deluxetable}


\section{Setting the photometric reference system}\label{sec:photo}
Photometry for our candidate spectrophotometric standards needs to be placed on a common flux scale at the top of the atmosphere. 
To achieve this goal we observed the three \emph{HST} primary CALSPEC WDs and the candidate DAWDs by using the same instrument and telescope set-up in Cycle 22. These observations allowed us to determine the instrumental zero-points (ZPs) for each WFC3 filter, and to tie the magnitudes of all the targets to the same photometric system.

As a first step we calculated fluxes and magnitudes in the AB photometric system for the  \emph{HST} primary 
CALSPEC WDs by using the HST tool {\it Pysynphot}\footnote{http://pysynphot.readthedocs.io/en/latest/using\_pysynphot.html}.
For these simulations, we used the latest model spectra of the three DAWDs provided by the CALSPEC database
({\it mod\_010}), which are calculated with the 
Non-Local Thermal-Equilibrium (NLTE) code from \citet{rauch2013}. These 
models are normalized to an absolute flux level defined by the flux of 3.44$\times$10$^{-9}$ 
erg cm$^{-2}$ s$^{-1}$ \AA$^{-1}$ for Vega at 0.5556$\mu$m, as reconciled with the MSX mid-IR 
absolute flux measures (B14).

The AB magnitude system \citep{oke74} is strictly speaking defined for monochromatic fluxes. 
If the flux at frequency $\nu$ is denoted by $f_{\nu}$ and expressed
in units of erg cm$^{-2}$ s$^{-1}$ Hz$^{-1}$, the corresponding AB magnitude at $\nu$ is defined by:

\begin{equation}
m(AB_{\nu}) = -2.5 \log (f_{\nu}) - 48.60
\end{equation}

This corresponds to a normalization where an object with a flat spectrum has AB magnitude equal to its $V$ band magnitude \citep{oke1983}.

To incorporate the idea of AB magnitudes for non-monochromatic use, say for a passband $X$, we use the extension as proposed 
by \citet{fukugita1996} for a photon proportional detector system to define the quantity $f_{X}$:

\begin{equation}
f_{X} = \frac{\int f_{\nu} v^{-1}R d\nu }{\int \nu^{-1} R d\nu} = \frac{\int N_{\nu} R d\nu }{\int (h\nu)^{-1} R d\nu}
\end{equation}

where $R$ is the (telescope + instrument + filter) response function for passband $X$, $N_{\nu}$ is the count rate of photons per unit frequency and $h$ is Planck's constant. The numerator on rightmost side is the photon count rate in the band, so $f_{X}$ is directly proportional to the photon count rate. 

The AB magnitude for passband $X$ is then given by:

\begin{equation}
m(AB_{X}) = -2.5 \log(f_{X}) - 48.60
\end{equation}


A characteristic wavelength, {\it pivot wavelength}, is defined to transform flux densities from the frequency to the
wavelength domain as:

\begin{equation}
\lambda _{p} = \sqrt{\frac{c f_{\nu}}{f_{\lambda }}} = \sqrt{\frac{\int R \lambda d\lambda }{\int R \frac{d\lambda }{\lambda } } }
\end{equation}

which is a source independent quantity. 
The Space Telescope (ST) magnitude system is defined in the wavelength domain
for passband $X$ as:

\begin{equation}
m(ST_{X}) = -2.5 log (f_{X}) - 21.10
\end{equation}

where ST mag = 0 is 3.63$\times$10$^{-9}$ erg cm$^{-2}$ s$^{-1}$ \AA$^{-1}$.
Having defined  $\lambda _{p}$, we can then convert AB to ST magnitudes with the relation:

\begin{equation}
m(ST_{\lambda}) = m(AB_{\nu}) + 5 log( \lambda _{p}) - 18.70
\end{equation}

We used {\it Pysynphot} to calculate synthetic fluxes and magnitudes in the AB photometric system 
for the primary WDs. As a reference we used the most updated files available on the Space Telescope
database\footnote{http://www.stsci.edu/hst/observatory/crds/throughput.html}. 
These reference files give the transmission curves for every element in the
optical path of the (HST + WFC3 + filter) system. For a complete list of the reference files please see the linked web-page.
The AB fluxes and magnitudes obtained for the three \emph{HST} primary 
CALSPEC WDs are listed in Table~\ref{table:6}.

\begin{figure*}
\begin{center}
\includegraphics[height=0.8\textheight,width=0.82\textwidth]{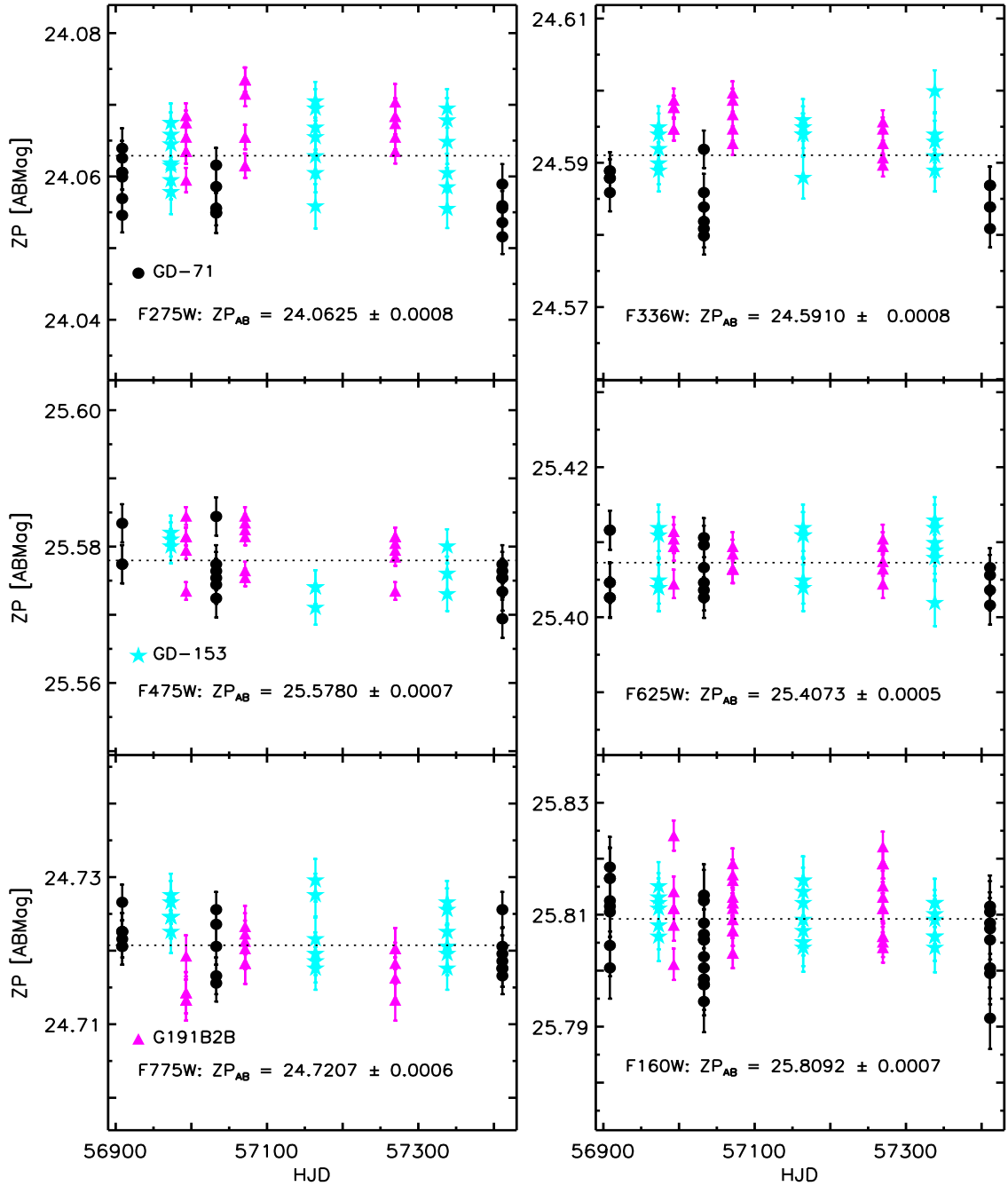}  
\caption{ZPs in the AB photometric system based on all the observations for the three \emph{HST} 
primary CALSPEC WDs (GD71 = black dots, GD153 = cyan stars, and G191B2B = magenta triangles) as 
a function of the Heliocentric Julian date (HJD) for six WFC3-UVIS and WFC3-IR filters as measured
with DAOPHOT. Error bars are shown and the derived ZPs are labeled. \label{fig:zp_all}}
\end{center}
\end{figure*}

The derived AB synthetic magnitudes are compared to the instrumental magnitudes 
measured from our observations in Cycle 22 for the primary WDs.
Fig.~\ref{fig:zp_all} shows the difference between synthetic and instrumental magnitudes as
a function of the observing epoch for the three stars. Observations were divided 
in multiple exposures for a total of three visits per star in a time interval of $\sim$ 1.3 year.
We performed a 1.5-$\sigma$ clipping on the data and we estimated the biweight mean of the difference
for the three primary WDs. This difference sets the ZP for our observations.
The estimated ZPs with their errors are labeled in each panel of Fig.~\ref{fig:zp_all}.
For observations in the $F275W$ and $F336W$ filters, GD71 
measurements (black dots) are consistently offset, i.e. fainter, compared to the other two WDs (cyan stars and
magenta triangles). The cause of this difference is not clear. 
A set of ZPs for each of the photometric reduction method was estimated and they are all 
listed in Table~\ref{table:7}.

As a sanity check we also derived ZPs for the same filters but for an aperture radius of 10 pixels for WFC3-UVIS, 
and to infinity for WFC3-UVIS and WFC3-IR, i.e. the aperture radii used by the WFC3 team to provide the official ZPs.
To derive ZPs to infinity we used the encircled energy (EE) correction tables provided by the 
WFC3 database\footnote{http://www.stsci.edu/hst/wfc3/analysis/ir\_ee; http://www.stsci.edu/hst/wfc3/analysis/uvis\_ee}.

\begin{figure*}
\begin{center}
\includegraphics[height=0.7\textheight,width=0.45\textwidth, angle=90]{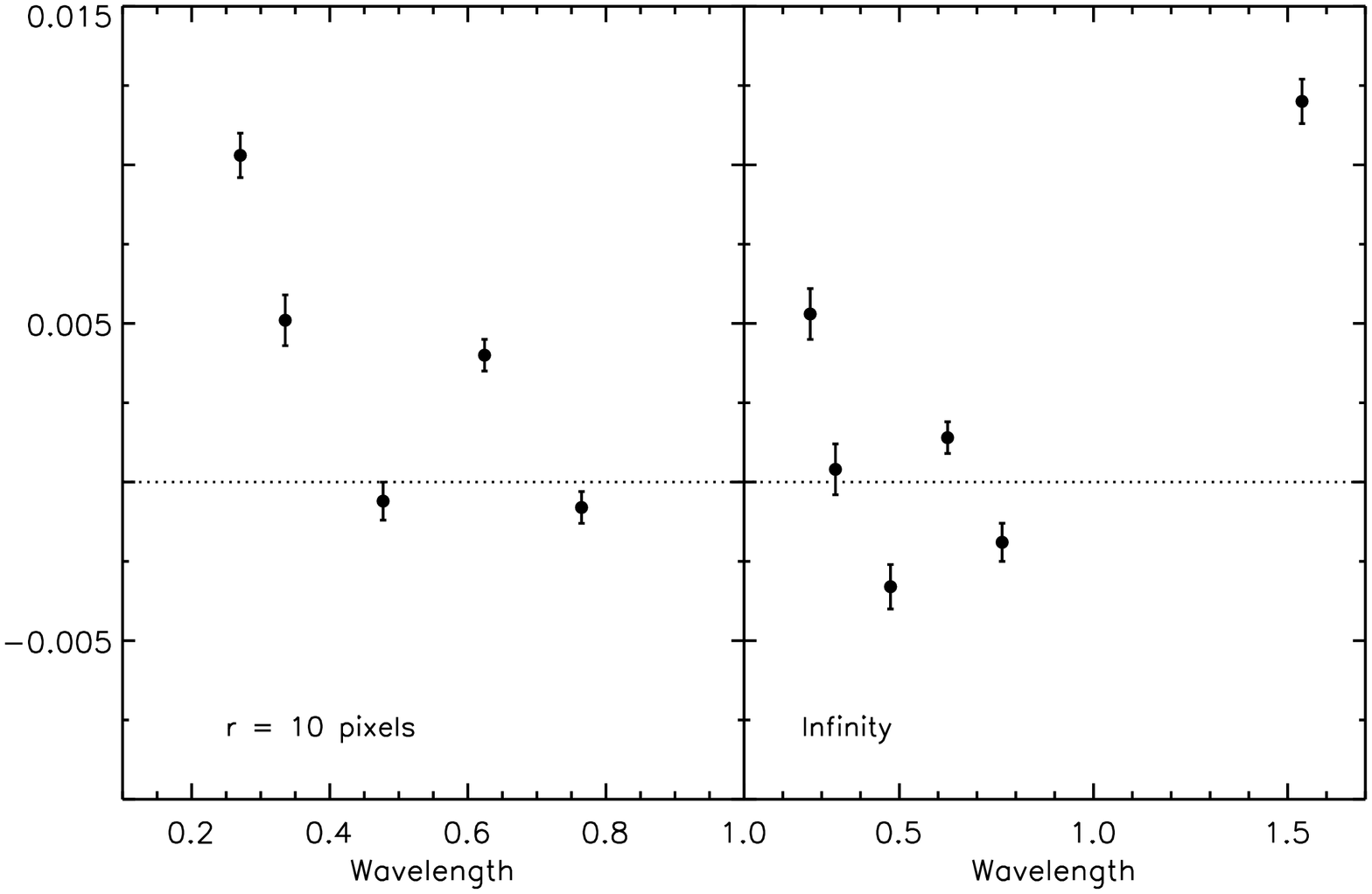}    
\caption{Comparison between ZPs in the AB photometric system measured from 
our observations of the three \emph{HST} primary CALSPEC WDs and the WFC3 official 
ZPs as a function of wavelength. 
ZPs are derived for an aperture radius of 10 pixels (left panel, UVIS) and infinity (right, UVIS + IR).
See text for more details. \label{fig:zp}}
\end{center}
\end{figure*}

Fig.~\ref{fig:zp} shows the comparison between WFC3 official ZPs and 
ZPs measured using our observations, reduced with DAOPHOT, for the three CALSPEC standards as a function of wavelength.
Error bars show uncertainties in our ZP estimates, since
there are no errors provided for the WFC3 ZPs.
The left panel shows the comparison for ZPs derived for an aperture radius of 10 pixels ($F160W$
is excluded since ZPs for WFC3-IR are not provided for this aperture), while the right panel
shows the same comparison for all filters and for an infinite aperture radius. 
The two sets of ZPs agree very well, with only the $F275W$ and the $F160W$ filters being $\gtrsim$1\% off. 
WFC3 official ZPs are calculated by using a set of observations taken
in between 2009 and 2015, and the epoch to which these sensitivities are normalized is then 
$\approx$ 2012.5, and they are an average of measurements collected on the UVIS2 amplifiers C and D.
The WFC3 inverse sensitivities change with time and our ZPs are provided for 
the average epoch of the observations, i.e. $\approx$ 2015.5, and are based on data collected
only on amplifier C. The change in sensitivity of the (detector+filter) system will be analyzed 
in Section~\ref{sec:time}. 
In spite of the aforementioned issues, the overall average difference between 
the two sets of ZP is 0.003 mag with a dispersion of 0.005 mag for a 10 pixel aperture, 
and 0.002 mag with a dispersion of 0.006 mag for the infinite aperture.
ZPs for a 10 pixel aperture radius and for infinity are listed in Table~\ref{table:8}.
These ZPs can be used by any astronomer performing observations by using WFC3-UVIS2
and WFC3-IR and to tie their photometry to the HST photometric scale.

\subsection{Tracking WFC3 sensitivity variation with time}\label{sec:time}
We used observations of the three \emph{HST} primary CALSPEC standards to track 
the variation of WFC3 sensitivity as a function of filter and time. The observations of the CALSPEC 
stars span a time interval of approximately 1.3 year, from the fall of 2014 to the beginning of 2016.
Instrumental count rates for aperture radii 7.5 (WFC3-UVIS) and 5 (WFC3-IR) pixels 
in the AB photometric system were derived for the CALSPEC WDs in the six filters for the three different visits, 
each one with a number of observations ranging from 6 to 8, 
depending on the filter, as described in section \ref{sec:obs}.
Synthetic count rates were derived with {\it Pysynphot} for the same stars as observed with WFC3.
We followed the same procedure described in section \ref{sec:photo} and we simulated
count rates for aperture radii of 7.5 (WFC3-UVIS) and 5 (WFC3-IR) pixels, i.e. the radii we used to perform 
photometry on the real images.

\begin{figure*}
\begin{center}
\includegraphics[height=0.85\textheight,width=0.8\textwidth]{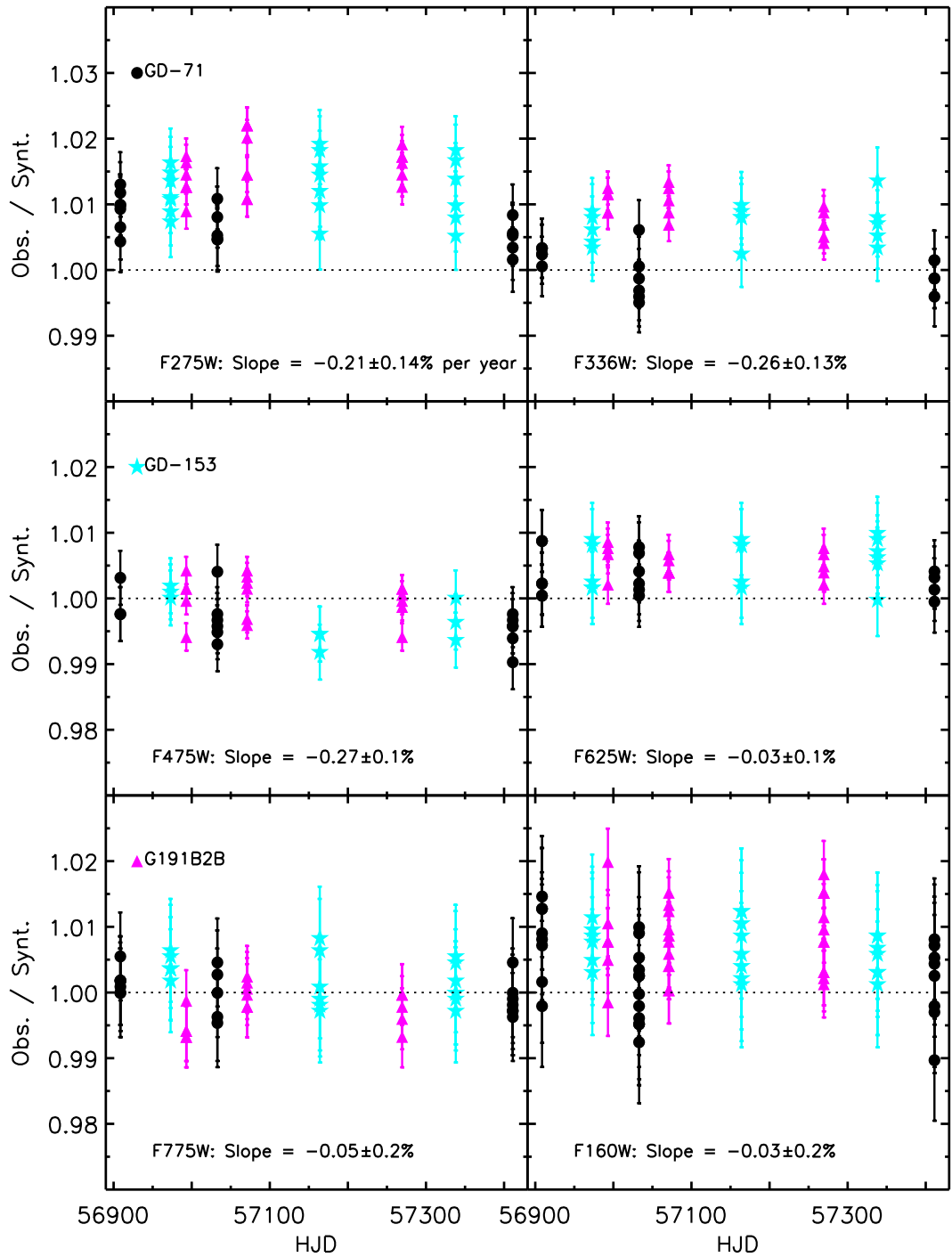}   
\caption{Ratio of observed to synthetic count rates for the three \emph{HST} primary CALSPEC standards 
(GD71 = black dots, GD153 = cyan stars, and G191B2B = magenta triangles) as a 
function of the Heliocentric Julian date (HJD) 
for six WFC3-UVIS and WFC3-IR filters. Error bars and the slope fits are shown.
The rate of yearly sensitivity loss is labeled. \label{fig:time}}
\end{center}
\end{figure*}

Fig.~\ref{fig:time} shows the ratio of the observed to synthetic count rates 
as a function of the observing epoch for the three primary WDs (GD71 = black dots, GD153 = cyan stars, and G191B2B = magenta triangles) and the six filters, after
we performed a 1.5-$\sigma$ clipping of the data.
The plot shows that the sensitivity is decreasing with time for all filters: 
the decrease is steeper for the bluer filters, $F275W$, $F336W$, and 
$F475W$, and shallower for the redder, $F625W,$ $F775W$, and $F160W$. 
We performed a linear least-square fit and obtained slopes ranging from -0.03 to -0.27, 
with the larger slopes for $F336W$ and $F475W$ and the smaller for $F625W$ 
and $F160W$. The fit to the data and the final sensitivity decrease rate per year are shown in Fig.~\ref{fig:time}. 
%

The sensitivity loss rates we obtained from our observations are in good agreement, within uncertainties, with the rates provided by the WFC3 photometric contamination monitor studies. 
One of the contamination monitor program is based on about 8 years of observations 
of the CALSPEC WD GWR70. These data show that WFC3 sensitivity decreases by less than 0.01\% for the 
ultraviolet filters $F275W$ and $F336W$ (see Table~ref{table:4} of \citealt{shanahan2017} for more details). 
However, no measurements are available for the $F475W$, $F625W$ and $F775W$ filters from this contamination monitor.
It is worth mentioning that UV filters had an increase in sensitivity soon after WFC3 was installed and then 
started to decrease (see Fig.~8 in \citealt{shanahan2017}). The very low percentage decrease obtained by \citet{shanahan2017} 
for the UV filters is due to fitting all the measurements for GWR70 at the same time.
A more recent contamination monitor study from the WFC3 team based on 8 years of photometry for 
the three primary WDs and the CALSPEC G-type standard P330E, obtained steeper slopes for the UV filters, 
by only considering measurements from when the sensitivity started to decrease. The new sensitivity loss rates range from $\sim$ -0.05 to -0.2\% per year for the UVIS filters $F275W$, $F336W$, $F475W$, $F625W$ and $F775W$ (private communication). 
These results will be soon published in a WFC3 ISR.

The WFC3 sensitivity loss rates that we derived by using our observations of the primary WDs 
have very large errors, 0.1-0.2\%. Our data cover indeed a very short time interval of a little more than 1 year and are insufficient to fully characterize the sensitivity variations with time.
On the other hand, the total dispersion of the measurements for the three primary WDs is always less 
than $\sim$ 0.005 mag in all UVIS filters and less then 1\% for in the infrared in the considered time interval
of our program observations.
Therefore, we did not apply any time correction to the photometry.

\subsection{The final magnitudes}\label{sec:mag}
The ZPs obtained in section~\ref{sec:photo} were applied to the weighted mean instrumental 
magnitudes of all the 23 candidate standard DAWDs. The ZPs were derived by using observations 
of the three \emph{HST} primary CALSPEC standards performed under the same conditions 
and reduced with the same technique and take into account any possible systematics in the observations and
data reduction process.
For each of the three different reduction methods, final calibrated magnitudes for filter $X$ in the 
AB photometric system are derived as:

\begin{equation}
m(AB_{X})_{cal} = m(AB_{X})_{inst} + ZP_{AB_X} = (-2.5 log(f_{X}) - 48.60) + ZP_{AB_X}
\end{equation}

where $f_X$ is in electrons/s, for aperture radii of 7.5 (WFC3-UVIS) and 5 pixels (WFC3-IR), respectively,
and $ZP_{AB_X}$ are the estimated ZPs listes in Table~\ref{table:7}.

The final magnitudes for the 23 candidate spectrophotometric standard DAWDs and the \emph{HST} primary 
CALSPEC WDs are listed in Table~\ref{table:9}.
Fig.~\ref{fig:dao_gaia} shows the $Bp - Rp, \ F475W - F775W$ color-color diagram for the 23
candidate spectrophotometric standard DAWDs where GAIA and WFC3 magnitudes derived with DAOPHOT are plotted.

Regardless of the photometric reduction method used, magnitudes for our DAWDs have an average dispersion 
ranging from 1 to 3 milli-mag for the WFC3-UVIS filters and from 5 to 10 milli-mag for the $F160W$ IR filter.

\clearpage
\startlongtable
\begin{deluxetable*}{lcccccccccccc}
\tabletypesize{\scriptsize}
\tablecaption{Photometry in the WFC3 UVIS and IR filters for the 3 \emph{HST} primary CALSPEC standards and the 23 candidate standard DAWDs 
in the AB photometric system. 
Photometry performed with three different software packages, DAOPHOT, Source Extractor and ILAPH and the
applied magnitude offsets between \emph{HST} Cycle 20 and Cycle 22, derived from Fig.~\ref{fig:offset}, are listed. 
See text for more details. \label{table:9}}
\tablehead{
\colhead{Star}&
\colhead{$F275W$}&
\colhead{$dF275W$}&
\colhead{$F336W$}& 
\colhead{$dF336W$}& 
\colhead{$F475W$}& 
\colhead{$dF475W$}& 
\colhead{$F625W$}& 
\colhead{$dF625W$}&
\colhead{$F775W$}&
\colhead{$dF775W$}&
\colhead{$F160W$}&
\colhead{$dF160W$}\\
\colhead{}&
\colhead{(mag)}&
\colhead{(mag)}&
\colhead{(mag)}&
\colhead{(mag)}&
\colhead{(mag)}&
\colhead{(mag)}&
\colhead{(mag)}&
\colhead{(mag)}&
\colhead{(mag)}&
\colhead{(mag)}&
\colhead{(mag)}&
\colhead{(mag)}
}
\startdata
\multicolumn{13}{c}{DAOPHOT} \\
\hline
Offsets                         & \ldots  & \ldots   & -0.033  & 0.001 & -0.007 & 0.005 & -0.006 & 0.004 & 0.013  & 0.006  & -0.014 & 0.004 \\
\hline
G191B2B 	                    & 10.488 & 0.002 & 10.888 & 0.001 & 11.498 & 0.001 & 12.030 & 0.001 & 12.451 & 0.001 & 13.883 & 0.002 \\ 
GD71		            & 11.986 & 0.002 & 12.333 & 0.001 & 12.796 & 0.001 & 13.277 & 0.001 & 13.672 & 0.001 & 15.065 & 0.002 \\ 
GD153		            & 12.199 & 0.002 & 12.565 & 0.001 & 13.099 & 0.002 & 13.597 & 0.001 & 14.002 & 0.001 & 15.413 & 0.002 \\ 
SDSSJ010322.19-002047.7    & 18.191 & 0.004 & 18.524 & 0.006 & 19.082 & 0.005 & 19.562 & 0.005 & 19.967 & 0.005 & 21.364 & 0.020 \\ 
SDSSJ022817.16-082716.4    & 19.512 & 0.006 & 19.732 & 0.037 & 19.811 & 0.005 & 20.178 & 0.006 & 20.506 & 0.007 & 21.737 & 0.015 \\ 
SDSSJ024854.96+334548.3    & 17.829 & 0.004 & 18.042 & 0.004 & 18.367 & 0.003 & 18.745 & 0.002 & 19.078 & 0.002 & 20.341 & 0.006 \\ 
SDSSJ041053.632-063027.580 & 18.110 & 0.009 & 18.401 & 0.004 & 18.879 & 0.004 & 19.254 & 0.003 & 19.387 & 0.007 & 19.500 & 0.005 \\ 
WD0554-165		   & 16.774 & 0.005 & 17.150 & 0.003 & 17.720 & 0.005 & 18.221 & 0.002 & 18.622 & 0.007 & 20.046 & 0.002 \\ 
SDSSJ072752.76+321416.1    & 17.158 & 0.003 & 17.467 & 0.003 & 17.990 & 0.003 & 18.456 & 0.002 & 18.839 & 0.002 & 20.214 & 0.006 \\ 
SDSSJ081508.78+073145.7    & 18.939 & 0.005 & 19.262 & 0.006 & 19.713 & 0.004 & 20.186 & 0.004 & 20.578 & 0.005 & 21.967 & 0.015 \\ 
SDSSJ102430.93-003207.0    & 18.248 & 0.038 & 18.509 & 0.004 & 18.903 & 0.004 & 19.314 & 0.005 & 19.667 & 0.009 & 20.989 & 0.014 \\ 
SDSSJ111059.42-170954.2    & 17.039 & 0.004 & 17.351 & 0.004 & 17.864 & 0.002 & 18.313 & 0.002 & 18.690 & 0.002 & 20.057 & 0.005 \\ 
SDSSJ111127.30+395628.0    & 17.432 & 0.004 & 17.832 & 0.005 & 18.419 & 0.003 & 18.940 & 0.004 & 19.344 & 0.002 & 20.795 & 0.010 \\ 
SDSSJ120650.504+020143.810 & 18.236 & 0.004 & 18.484 & 0.004 & 18.669 & 0.004 & 19.058 & 0.004 & 19.411 & 0.005 & 20.700 & 0.011 \\ 
SDSSJ121405.11+453818.5    & 16.938 & 0.002 & 17.279 & 0.002 & 17.758 & 0.002 & 18.231 & 0.002 & 18.630 & 0.002 & 20.035 & 0.004 \\ 
SDSSJ130234.43+101238.9    & 16.185 & 0.002 & 16.519 & 0.002 & 17.033 & 0.002 & 17.512 & 0.002 & 17.904 & 0.001 & 19.302 & 0.004 \\ 
SDSSJ131445.050-031415.588 & 18.254 & 0.004 & 18.593 & 0.004 & 19.100 & 0.004 & 19.571 & 0.004 & 19.933 & 0.010 & 21.329 & 0.012 \\ 
SDSSJ151421.27+004752.8    & 15.108 & 0.002 & 15.387 & 0.002 & 15.707 & 0.002 & 16.119 & 0.001 & 16.470 & 0.001 & 17.783 & 0.004 \\ 
SDSSJ155745.40+554609.7    & 16.496 & 0.002 & 16.873 & 0.002 & 17.468 & 0.003 & 17.990 & 0.002 & 18.389 & 0.002 & 19.832 & 0.005 \\ 
SDSSJ163800.360+004717.822 & 18.012 & 0.007 & 18.314 & 0.004 & 18.838 & 0.004 & 19.283 & 0.003 & 19.664 & 0.005 & 20.999 & 0.015 \\ 
SDSSJ172135.97+294016.0    & 20.370 & 0.010 & 20.086 & 0.014 & 19.654 & 0.004 & 19.670 & 0.003 & 19.769 & 0.003 & 20.554 & 0.022 \\ 
SDSSJ181424.075+785403.048 & 15.788 & 0.002 & 16.119 & 0.002 & 16.542 & 0.002 & 17.004 & 0.002 & 17.392 & 0.001 & 18.782 & 0.002 \\ 
SDSSJ203722.169-051302.964 & 18.254 & 0.007 & 18.540 & 0.004 & 18.940 & 0.006 & 19.371 & 0.007 & 19.674 & 0.008 & 20.965 & 0.009 \\ 
SDSSJ210150.65-054550.9    & 18.064 & 0.003 & 18.328 & 0.004 & 18.654 & 0.003 & 19.062 & 0.002 & 19.419 & 0.003 & 20.737 & 0.006 \\ 
SDSSJ232941.330+001107.755 & 17.940 & 0.003 & 18.105 & 0.004 & 18.158 & 0.005 & 18.472 & 0.003 & 18.785 & 0.006 & 19.997 & 0.007 \\ 
SDSSJ235144.29+375542.6    & 17.446 & 0.003 & 17.658 & 0.002 & 18.073 & 0.002 & 18.459 & 0.002 & 18.788 & 0.002 & 20.070 & 0.004 \\ 
\hline
\multicolumn{13}{c}{SExtractor} \\
\hline
Offsets                         & \ldots  & \ldots   & -0.031  & 0.003 & -0.004 & 0.003 & -0.009 & 0.002 & 0.012  & 0.004  & -0.011 & 0.005 \\
\hline
G191B2B 	           & 10.488 & 0.002 & 10.889 & 0.001 & 11.497 & 0.001 & 12.029 & 0.001 & 12.451 & 0.001 & 13.884 & 0.001 \\ 
GD71		           & 11.989 & 0.002 & 12.335 & 0.001 & 12.798 & 0.001 & 13.277 & 0.001 & 13.671 & 0.001 & 15.063 & 0.002 \\ 
GD153		           & 12.196 & 0.002 & 12.562 & 0.001 & 13.097 & 0.002 & 13.598 & 0.001 & 14.003 & 0.001 & 15.414 & 0.002 \\ 
SDSSJ010322.19-002047.7    & 18.197 & 0.004 & 18.532 & 0.005 & 19.088 & 0.005 & 19.564 & 0.004 & 19.968 & 0.005 & 21.359 & 0.010 \\ 
SDSSJ022817.16-082716.4    & 19.531 & 0.006 & 19.761 & 0.018 & 19.823 & 0.006 & 20.178 & 0.005 & 20.514 & 0.004 & 21.740 & 0.013 \\ 
SDSSJ024854.96+334548.3    & 17.840 & 0.003 & 18.047 & 0.004 & 18.368 & 0.003 & 18.745 & 0.002 & 19.079 & 0.002 & 20.335 & 0.005 \\ 
SDSSJ041053.632-063027.580 & 18.109 & 0.009 & 18.410 & 0.004 & 18.884 & 0.004 & 19.256 & 0.003 & 19.387 & 0.005 & 19.500 & 0.005 \\ 
WD0554-165		   & 16.777 & 0.005 & 17.153 & 0.004 & 17.729 & 0.003 & 18.222 & 0.003 & 18.619 & 0.005 & 20.043 & 0.006 \\ 
SDSSJ072752.76+321416.1    & 17.164 & 0.003 & 17.474 & 0.003 & 17.993 & 0.002 & 18.457 & 0.002 & 18.840 & 0.002 & 20.214 & 0.005 \\ 
SDSSJ081508.78+073145.7    & 18.965 & 0.007 & 19.280 & 0.005 & 19.714 & 0.005 & 20.185 & 0.004 & 20.579 & 0.005 & 21.967 & 0.012 \\ 
SDSSJ102430.93-003207.0    & 18.264 & 0.014 & 18.517 & 0.004 & 18.909 & 0.004 & 19.314 & 0.003 & 19.668 & 0.008 & 20.994 & 0.010 \\ 
SDSSJ111059.42-170954.2    & 17.047 & 0.003 & 17.359 & 0.004 & 17.867 & 0.002 & 18.314 & 0.002 & 18.689 & 0.002 & 20.054 & 0.005 \\ 
SDSSJ111127.30+395628.0    & 17.437 & 0.004 & 17.838 & 0.005 & 18.424 & 0.003 & 18.940 & 0.004 & 19.346 & 0.002 & 20.790 & 0.008 \\ 
SDSSJ120650.504+020143.810 & 18.244 & 0.004 & 18.491 & 0.004 & 18.672 & 0.004 & 19.060 & 0.003 & 19.412 & 0.004 & 20.703 & 0.006 \\ 
SDSSJ121405.11+453818.5    & 16.943 & 0.003 & 17.283 & 0.002 & 17.759 & 0.002 & 18.231 & 0.002 & 18.631 & 0.002 & 20.036 & 0.004 \\ 
SDSSJ130234.43+101238.9    & 16.191 & 0.002 & 16.522 & 0.002 & 17.036 & 0.002 & 17.514 & 0.002 & 17.903 & 0.001 & 19.302 & 0.004 \\ 
SDSSJ131445.050-031415.588 & 18.258 & 0.004 & 18.600 & 0.005 & 19.106 & 0.004 & 19.572 & 0.004 & 19.936 & 0.008 & 21.327 & 0.008 \\ 
SDSSJ151421.27+004752.8    & 15.112 & 0.002 & 15.390 & 0.002 & 15.708 & 0.002 & 16.119 & 0.002 & 16.470 & 0.001 & 17.783 & 0.004 \\ 
SDSSJ155745.40+554609.7    & 16.501 & 0.002 & 16.877 & 0.002 & 17.472 & 0.002 & 17.991 & 0.002 & 18.389 & 0.002 & 19.832 & 0.005 \\ 
SDSSJ163800.360+004717.822 & 18.013 & 0.007 & 18.321 & 0.004 & 18.842 & 0.004 & 19.285 & 0.003 & 19.663 & 0.004 & 21.002 & 0.007 \\ 
SDSSJ172135.97+294016.0    & 20.374 & 0.013 & 20.089 & 0.012 & 19.666 & 0.005 & 19.670 & 0.003 & 19.769 & 0.003 & 20.551 & 0.024 \\ 
SDSSJ181424.075+785403.048 & 15.792 & 0.002 & 16.122 & 0.002 & 16.543 & 0.002 & 17.005 & 0.002 & 17.392 & 0.001 & 18.783 & 0.002 \\ 
SDSSJ203722.169-051302.964 & 18.262 & 0.004 & 18.549 & 0.004 & 18.947 & 0.005 & 19.377 & 0.003 & 19.677 & 0.005 & 20.981 & 0.020 \\ 
SDSSJ210150.65-054550.9    & 18.075 & 0.003 & 18.337 & 0.003 & 18.655 & 0.003 & 19.063 & 0.002 & 19.419 & 0.003 & 20.741 & 0.006 \\ 
SDSSJ232941.330+001107.755 & 17.947 & 0.004 & 18.110 & 0.004 & 18.161 & 0.004 & 18.473 & 0.003 & 18.784 & 0.004 & 19.993 & 0.006 \\ 
SDSSJ235144.29+375542.6    & 17.456 & 0.003 & 17.666 & 0.003 & 18.074 & 0.002 & 18.461 & 0.002 & 18.790 & 0.002 & 20.069 & 0.004 \\ 
\hline
\multicolumn{13}{c}{ILAPH} \\
\hline
Offsets                         & \ldots  & \ldots   & -0.033  & 0.003 & -0.009 & 0.004 & -0.014 & 0.002 & 0.009  & 0.004  & -0.012 & 0.005 \\
\hline
G191B2B 	                 & 10.490 & 0.001 & 10.890 & 0.001 & 11.499 & 0.001 & 12.031 & 0.001 & 12.451 & 0.001 & 13.885 & 0.002 \\ 
GD71		           & 11.989 & 0.001 & 12.336 & 0.001 & 12.799 & 0.001 & 13.279 & 0.001 & 13.672 & 0.001 & 15.068 & 0.002 \\ 
GD153		           & 12.201 & 0.002 & 12.568 & 0.001 & 13.100 & 0.002 & 13.598 & 0.001 & 14.002 & 0.001 & 15.414 & 0.002 \\ 
SDSSJ010322.19-002047.7    & 18.195 & 0.004 & 18.527 & 0.005 & 19.083 & 0.005 & 19.569 & 0.005 & 19.965 & 0.006 & 21.355 & 0.012 \\ 
SDSSJ022817.16-082716.4    & 19.518 & 0.008 & 19.715 & 0.010 & 19.815 & 0.007 & 20.169 & 0.007 & 20.501 & 0.006 & 21.737 & 0.017 \\ 
SDSSJ024854.96+334548.3    & 17.828 & 0.003 & 18.040 & 0.006 & 18.370 & 0.003 & 18.746 & 0.002 & 19.077 & 0.002 & 20.340 & 0.006 \\ 
SDSSJ041053.632-063027.580 & 18.116 & 0.009 & 18.404 & 0.004 & 18.879 & 0.005 & 19.254 & 0.003 & 19.393 & 0.005 & 19.498 & 0.005 \\ 
WD0554-165		   & 16.776 & 0.005 & 17.153 & 0.003 & 17.727 & 0.005 & 18.220 & 0.002 & 18.617 & 0.005 & 20.043 & 0.007 \\ 
SDSSJ072752.76+321416.1    & 17.163 & 0.003 & 17.471 & 0.003 & 17.993 & 0.003 & 18.457 & 0.002 & 18.837 & 0.003 & 20.217 & 0.007 \\ 
SDSSJ081508.78+073145.7    & 18.950 & 0.006 & 19.263 & 0.008 & 19.716 & 0.005 & 20.184 & 0.005 & 20.579 & 0.006 & 21.962 & 0.024 \\ 
SDSSJ102430.93-003207.0    & 18.261 & 0.018 & 18.514 & 0.004 & 18.904 & 0.004 & 19.317 & 0.004 & 19.665 & 0.010 & 20.990 & 0.013 \\ 
SDSSJ111059.42-170954.2    & 17.041 & 0.003 & 17.354 & 0.004 & 17.867 & 0.003 & 18.313 & 0.002 & 18.689 & 0.002 & 20.057 & 0.005 \\ 
SDSSJ111127.30+395628.0    & 17.443 & 0.004 & 17.830 & 0.006 & 18.420 & 0.003 & 18.939 & 0.004 & 19.344 & 0.002 & 20.797 & 0.009 \\ 
SDSSJ120650.504+020143.810 & 18.240 & 0.004 & 18.489 & 0.004 & 18.672 & 0.004 & 19.060 & 0.003 & 19.411 & 0.007 & 20.703 & 0.008 \\ 
SDSSJ121405.11+453818.5    & 16.940 & 0.002 & 17.283 & 0.002 & 17.761 & 0.002 & 18.236 & 0.003 & 18.629 & 0.002 & 20.038 & 0.004 \\ 
SDSSJ130234.43+101238.9    & 16.188 & 0.002 & 16.522 & 0.002 & 17.036 & 0.002 & 17.514 & 0.002 & 17.904 & 0.002 & 19.303 & 0.004 \\ 
SDSSJ131445.050-031415.588 & 18.258 & 0.004 & 18.597 & 0.005 & 19.102 & 0.005 & 19.567 & 0.005 & 19.955 & 0.009 & 21.328 & 0.012 \\ 
SDSSJ151421.27+004752.8    & 15.110 & 0.002 & 15.391 & 0.002 & 15.709 & 0.002 & 16.120 & 0.002 & 16.471 & 0.001 & 17.787 & 0.004 \\ 
SDSSJ155745.40+554609.7    & 16.500 & 0.002 & 16.877 & 0.002 & 17.470 & 0.003 & 17.992 & 0.002 & 18.388 & 0.002 & 19.834 & 0.005 \\ 
SDSSJ163800.360+004717.822 & 18.016 & 0.007 & 18.318 & 0.004 & 18.840 & 0.005 & 19.281 & 0.003 & 19.660 & 0.005 & 20.996 & 0.009 \\ 
SDSSJ172135.97+294016.0      & 20.371 & 0.013 & 20.078 & 0.015 & 19.656 & 0.004 & 19.670 & 0.003 & 19.768 & 0.003 & 20.552 & 0.021 \\ 
SDSSJ181424.075+785403.048 & 15.791 & 0.002 & 16.121 & 0.002 & 16.544 & 0.002 & 17.001 & 0.002 & 17.393 & 0.001 & 18.786 & 0.002 \\ 
SDSSJ203722.169-051302.964 & 18.257 & 0.007 & 18.544 & 0.004 & 18.943 & 0.006 & 19.350 & 0.012 & 19.672 & 0.010 & 20.979 & 0.023 \\ 
SDSSJ210150.65-054550.9      & 18.068 & 0.004 & 18.334 & 0.004 & 18.656 & 0.003 & 19.064 & 0.002 & 19.414 & 0.004 & 20.740 & 0.008 \\ 
SDSSJ232941.330+001107.755 & 17.943 & 0.004 & 18.109 & 0.004 & 18.161 & 0.006 & 18.470 & 0.003 & 18.775 & 0.007 & 19.995 & 0.006 \\ 
SDSSJ235144.29+375542.6     & 17.449 & 0.004 & 17.662 & 0.003 & 18.075 & 0.003 & 18.459 & 0.003 & 18.787 & 0.002 & 20.075 & 0.004 \\ 
\enddata
\end{deluxetable*}


\section{Summary and conclusions}
In this paper we presented the methods used to provide sub-percent precision photometry for a set of faint candidate 
spectrophotometric standard DAWDs. We also presented data reduction processes, and possible source of uncertainties,
of spectroscopic data collected for the same stars. These spectra are used to derive temperature and 
surface gravities for the candidate standards. 

In order to investigate the possible sources of systematics and to derive reliable uncertainties for the DAWD photometry, 
we used three different software packages to reduce WFC3 data, DAOPHOT, ILAPH, and SExtractor. 
Our analysis showed that photometry performed with the first two packages agrees very well within uncertainties, 
while photometry from SExtractor shows larger dispersion and a trend for which magnitudes of fainter stars results to
be fainter compared to DAOPHOT and ILAPH magnitudes, in particular in the bluest filters, $F275W$ and $F336W$. 
This trend is probably due to SExtractor over-estimating the sky background.

We tested our data for photometric uncertainties due to the presence of 
external or internal persistence on the IR images. 
We found that the largest fraction of pixels affected by a persistence signal higher than 0.01 e$^{-}$/s is 0.33\%, 
for images of star SDSSJ181424.075+785403.048.
However, the affected pixels do not overlap with the location of the star on the images. 
Our observing strategy was devised to avoid self-persistence in our exposures.

CRNL in WFC3-IR exposures was estimated to be 0.010$\pm$0.0025 mag 
per dex, and might slightly affect our observations. However, 
we do not apply any CRNL corrections on the photometry presented here and
we plan to fully characterize this effect in NA19.

Our data show no systematics in the photometry due to the WFC3-UVIS shutter shading effect
for an aperture radius of 7.5 pixels.
We tested photometry on 1s exposures for G191B2B and we found that the dispersion of the measurements on images collected 
by using shutter blade B, $\sigma$ = 0.004 mag, is about the same compared to the 
dispersion on images observed with blade A, $\sigma$ = 0.005 mag.
All the other DAWDs were observed with exposure times $\gtrsim$ 5s, and so observations
are not affected by shutter shading. 

The presence of unseen companion stars could also introduce uncertainties/systematics in the photometry.
However, the observed DAWDs are all in very sparse fields, with a maximum of other 3 objects including the DAWD 
in the $\sim$ 20$\times$20\arcsec~ observed field of view down to $g \sim$ 23 mag. 
Our simulations also showed that stars fainter than 6 mag compared to the target WDs cannot affect the 
photometry of the target DAWD even if falling inside the aperture radius. We can then safely assume
that the photometry of our set of standards cannot be contaminated by unseen neighbor stars. 

Time-series observations collected with the LCO network of telescopes showed that most of our candidate spectrophotometric 
standards are stable. Two of them, namely SDSSJ20372.169-051302.964 and WD0554-165, show clear sign of variability in their light curves. 
The first star also show emission features in the Balmer lines of the spectra implying the presence of a low-mass companion. 
We do not know the origin of the variability for WD0554-165. Two other DAWDs, SDSSJ010322.10-002047.7 and SDSSJ102430.93-003207.0
show hints of variability, but these results need to be confirmed with further data. SDSSJ20372.169-051302.964 and WD0554-165 will be excluded by our set of candidate standard DAWDs.

We used observations of the three \emph{HST} primary CALSPEC standards, collected at the same time 
as our target DAWDs, to estimate ZPs in the AB photometric system to be applied to instrumental magnitudes of all the observed targets. 

We also derived ZPs in the AB photometric system for six WFC3 filters, namely $F275W$, $F336W$,
$F475W$, $F625W$, $F775W$ and $F160W$, for a 10 pixel aperture radius and for infinity.
The ZPs are provided in Table~\ref{table:8} and can be used to calibrate any WFC3-UVIS2 photometry.

We also verified for the presence of WFC3 sensitivity changes during the $\sim$ 1.3 years of the observations by using the same data. 
A decrease in sensitivity is observed in all six filters, with the largest percentage decline in sensitivity for $F475W$ (-0.27\%) and the 
smallest for $F625W$ and $F160W$ (-0.03\%), but our data do not span a sufficient time interval to fully characterize WFC3
sensitivity behavior. However, the  overall dispersion of the measurements over the time interval of our observations 
is less than 0.5\% for WFC3-UVIS
and less than 1\% for WFC3-IR. Therefore, we do not apply any time correction to our photometry. 

We provided final calibrated AB magnitudes in five WFC3-UVIS filters and one IR filter for the 23 candidate spectrophotometric 
standard DAWDs and the three \emph{HST} primary CALSPEC standards obtained by using the three different software packages,
DAOPHOT, SExtractor and ILAPH.
Magnitudes have an average dispersion in the range 1 to 3 milli-mag for WFC3-UVIS filters and 5 to 10 milli-mag for the $F160W$ IR filter. 
Photometry is available as machine readable Table~\ref{table:9}.

Synthetic magnitudes in different photometric systems, such as PS, GAIA, and SDSS, for the set of standard DAWDs will be calculated and 
provided in NA19.


\acknowledgments{
{\it Acknowledgments:}
This study was supported by NASA through grants GO-12967 and GO-13711 from the National
Optical Astronomy Observatory, which is operated by AURA, Inc., and grant GO-15113 from
the Space Telescope Science Institute, which is operated by AURA, Inc., under 
NASA contract NAS 5-26555.
E. Olszewski was also partially supported by the NSF through grants AST- 1313006  and AST-1815767.
This work has made use of data from the European Space Agency (ESA) mission
{\it Gaia} (\url{https://www.cosmos.esa.int/gaia}), processed by the {\it Gaia}
Data Processing and Analysis Consortium (DPAC,
\url{https://www.cosmos.esa.int/web/gaia/dpac/consortium}). Funding for the DPAC
has been provided by national institutions, in particular the institutions
participating in the {\it Gaia} Multilateral Agreement.}

\facility{
\emph{HST} (WFC3), \emph{GEMINI} , \emph{GAIA}, \emph{Pan-STARRS}
}



\clearpage
\bibliographystyle{aa}

\bibliography{calamida}

\appendix

\startlongtable
\begin{deluxetable*}{lc}
\tablecaption{List of acronyms used in the manuscript grouped by class and in alphabetical order.  \label{table:10}}
\tablehead{
\colhead{Acronym}&
\colhead{Meaning}\\
\colhead{}&
\colhead{}
}
\startdata
\multicolumn{2}{c}{Instruments/Detectors} \\
\hline
ACS   & Advanced Camera for Surveys \\
GALEX & Galaxy Evolution Explorer  \\
GMOS  & Gemini Multi-Object Spectrograph \\
HST   & Hubble Space Telescope  \\
LCO   & Las Cumbres Observatory \\
LSST  & Large Synoptic Survey Telescope \\
MMT  & Multiple Mirror Telescope \\
STIS  & Space Telescope Imaging spectrograph  \\
UVIS1/2 & Chips of the Wide Field Camera 3 detector \\
WFC3 - IR   & Wide Field Camera 3 Infrared detector  \\
WFC3 - UVIS & Wide Field Camera 3 Ultraviolet and VISual detector  \\
WISE  & Wide-field Infrared Survey Explorer \\
\hline
\multicolumn{2}{c}{Surveys} \\
\hline
ASAS-SN & All Sky Automated Survey for SuperNovae  \\
ATLAS    & Asteroid Terrestrial-impact Last Alert System  \\
DES      & Dark Energy Survey  \\
PS        & Pan-STARSS  \\
SDSS    & Sloan Digital Sky Survey  \\
ZTF         & Zwicky Transient Factory \\
\hline
\multicolumn{2}{c}{Software packages} \\
\hline 
ALLFRAME & Routine to perform simultaneous point-spread function photometry on different images \citep{stetson1994} \\
DAOPHOT   & DAOPHOTIV group of routines to perform aperture and point-spread function photometry \citep{stetson1987} \\
Drizzle Pac  & Software to stack images collected with Hubble Space Telescope  \\
ILAPH         & IDL routines to perform aperture photometry from Abhijit Saha \\
SExtractor  & Source Extractor software to perform aperture and point-spread function photometry \citep{bertin1996} \\
\hline
\multicolumn{2}{c}{Others} \\
\hline 
AS         & Artificial star  \\
cal\_wf3   & Image calibration pipeline for the Wide Field Camera 3 detectors \\
CALSPEC & Database of the Hubble Space Telescope spectrophotometric standard stars  \\
CTE         & Charge Transfer Efficiency \\
CR           & Cosmic ray \\
CRNL      & Count-Rate Non-Linearity \\
DAWD    & Hydrogen atmosphere white dwarf \\
FoV           & Field of view \\
FLUXCORR & Image header keyword indicating if the flux scaling needs to be performed \\
FWHM      & Full-Width Half maximum \\
IMPHTTAB      & Image photometry reference table \\
IR-FIX              & Fixed Aperture centered on the Wide Field Camera 3 Infrared detector \\
IRSUB256-FIX & 256$\times$256 pixel sub-aperture on the center of the Wide Field Camera 3 Infrared detector \\
IRSUB512-FIX & 512$\times$512 pixel sub-aperture on the center of the Wide Field Camera 3 Infrared detector \\
MODTRAN       & MODerate resolution atmospheric TRANsmission \\
NIR               & Near-infrared  \\
PAM             & Pixel Area Map \\
PHOTFLAM  & Image header keyword for the inverse sensitivity \\
PSF              & Point-spread function \\
SED             & Spectral Energy Distribution  \\
UV               & Ultraviolet  \\
UVIS1-FIX   & Fixed Aperture centered on the UVIS1 chip of the Wide Field Camera 3 Ultraviolet and VISual detector \\
UVIS2-C512C-SUB & 512$\times$512 pixel sub-aperture on the corner of the UVIS2 chip of the Wide Field Camera 3 Ultraviolet and VISual detector \\
ZP                            &   Zero point  \\
\hline 
\hline 
\enddata
\end{deluxetable*}

\end{document}